\documentclass[11pt]{article}
\usepackage[preprint]{neurips_2025}
\usepackage[margin=1in]{geometry}

\usepackage{lmodern}
\usepackage{amsmath}
\usepackage{amssymb}
\usepackage{graphicx}
\usepackage{booktabs}
\usepackage{float}
\usepackage{hyperref}
\usepackage{xcolor}
\usepackage{listings}
\usepackage{natbib}
\usepackage{url}

\hypersetup{
  colorlinks=true,
  linkcolor=blue!70!black,
  citecolor=green!50!black,
  urlcolor=blue!60!black,
}

\lstdefinelanguage{yaml}{
  keywords={true, false, null},
  morecomment=[l]{\#},
  morestring=[b]',
  morestring=[b]",
  sensitive=true,
}

\title{Guess-Verify-Refine: Data-Aware Top-K for Sparse-Attention Decoding on Blackwell via Temporal Correlation}

\author{
  Long Cheng, Ritchie Zhao, Timmy Liu, Mindy Li, Xianjie Qiao, Kefeng Duan,\\
  Yu-Jung Chen, Xiaoming Chen, Bita Darvish Rouhani, and June Yang\thanks{Corresponding author: June Yang (\texttt{juney@nvidia.com}).}\\
  NVIDIA
}

\begin{document}

\maketitle

\begin{abstract}
Sparse-attention decoders rely on exact Top-K selection to choose the most important
key-value entries for each query token. In long-context LLM serving, this Top-K stage runs
once per decode query and becomes a meaningful latency bottleneck even when the indexer and
attention kernels are already highly optimized. We present \textbf{Guess-Verify-Refine
(GVR)}, a data-aware exact Top-K algorithm for sparse-attention decoding on NVIDIA
Blackwell. GVR exploits temporal correlation across consecutive decode steps: it uses the
previous step's Top-K as a prediction signal, computes pre-indexed statistics, narrows to a
valid threshold by secant-style counting in 1--2 global passes, verifies candidates with a
ballot-free collector, and finishes exact selection in shared memory. We connect this
behavior to the Toeplitz / RoPE structure of DeepSeek Sparse Attention (DSA) indexer scores
and validate the design on real DeepSeek-V3.2 workloads integrated into TensorRT-LLM. GVR
achieves an average \textbf{1.88$\times$} single-operator speedup over the production
radix-select kernel, with up to \textbf{2.42$\times$} per layer per step, while preserving
bit-exact Top-K outputs. In controlled TEP8 min-latency deployment, it improves end-to-end
TPOT by up to \textbf{7.52\%} at 100K context, with larger gains at longer contexts and
smaller but still positive gains under speculative decoding. While implemented and validated
in the current TensorRT-LLM DSA stack on Blackwell, the same principle may extend to
sparse-attention decoders whose decode-phase Top-K exhibits temporal stability.
\end{abstract}

\section{Introduction}
\label{sec:introduction}

Agentic and code-centric LLM workloads routinely push decode-time context lengths into the
100K+ regime. Sparse attention removes the quadratic attention bottleneck, but it does not
remove the need to rank a long vector of indexer scores at every decode step. In practical
sparse-attention deployments such as DeepSeek
DSA~\citep{deepseekv32} and related systems~\citep{nsa,moba,rocketkv,quest,sagekv}, exact
\textbf{Top-K selection} remains on the critical path, and its latency grows with sequence
length even after the sparse MLA kernel and indexer MQA kernel have been highly optimized.

Prior GPU Top-K work has largely treated the operator as a \emph{distribution-agnostic}
primitive. Exact GPU algorithms based on radix decomposition, sampling, or bucketed
selection~\citep{radik,zois2019topk,zhangsc23} optimize memory traffic and parallelism across
wide $(N,K)$ regimes, while approximate variants trade exactness for additional
parallelism~\citep{approxtopk}. These methods are essential baselines, but they do not
exploit a signal that is specific to autoregressive sparse-attention decoding: consecutive
decode steps query highly similar neighborhoods of the KV cache. In inference, where exact
Top-K is required to preserve model behavior, this temporal stability creates an opportunity
for \emph{data-aware exact} selection rather than approximate pruning.

This paper develops that opportunity in the concrete setting of DeepSeek Sparse Attention
(DSA). DSA selects the top-2048 tokens from potentially tens or hundreds of thousands of
indexer scores via a lightweight \textbf{indexer} and \textbf{Top-K selector}. We introduce
\textbf{Guess-Verify-Refine (GVR)}, an exact Top-K algorithm that converts the previous
decode step's Top-K into a temporal prediction signal. GVR first computes pre-indexed
statistics, then uses secant-style threshold search to reduce full-row passes from 3--4 to
1--2, collects candidates with a ballot-free design, and completes exact selection in shared
memory. On real DeepSeek-V3.2 decoding workloads running on NVIDIA Blackwell GPUs, this
data-aware design achieves an average \textbf{1.88$\times$} single-operator speedup over the
production radix-select kernel---an evolution of \citet{zhangsc23} optimized for Blackwell by
the same team---with up to \textbf{2.42$\times$} per layer per step and no loss in output
accuracy.

Our contributions are threefold. First, we identify and formalize the temporal-correlation
signal behind decode-phase sparse-attention Top-K, connecting it to the Toeplitz / RoPE
structure of the indexer scores. Second, we design and implement a four-phase exact GPU
algorithm that turns that signal into fewer global-memory passes and lower synchronization
overhead. Third, we integrate the method into TensorRT-LLM~\citep{trtllm} and demonstrate
both kernel-level and end-to-end gains on real DeepSeek-V3.2 workloads, including up to
\textbf{7.52\%} TPOT reduction in TEP8 min-latency deployment. While demonstrated in the
current DSA implementation, the same principle may extend to sparse-attention decoders whose
Top-K exhibits temporal stability.

\section{Background: Indexer Top-K in DeepSeek Sparse Attention}
\label{sec:background}

\subsection{Lightweight Indexer and Top-K Selection}
\label{sec:lightweight-indexer}

As described in the DeepSeek-V3 technical report~\citep{deepseekv3}, DSA computes index
scores via a lightweight MQA mechanism:
\begin{equation}
  I_{t} = \sum_{j=1}^{h} W_j^I \cdot \mathrm{ReLU}\!\bigl(Q_{t,j}^I (K_t^I)^T\bigr)
  \label{eq:indexer}
\end{equation}
The index score tensor $I_t \in \mathbb{R}^N$ (where $N$ is the current sequence length)
quantifies the importance of each past key-value token for the current query token. A Top-K
operation then selects the $K = 2048$ highest-scoring positions, and only these are used for
the subsequent sparse MLA computation.

\begin{figure}[t]
  \centering
  \includegraphics[width=0.92\textwidth]{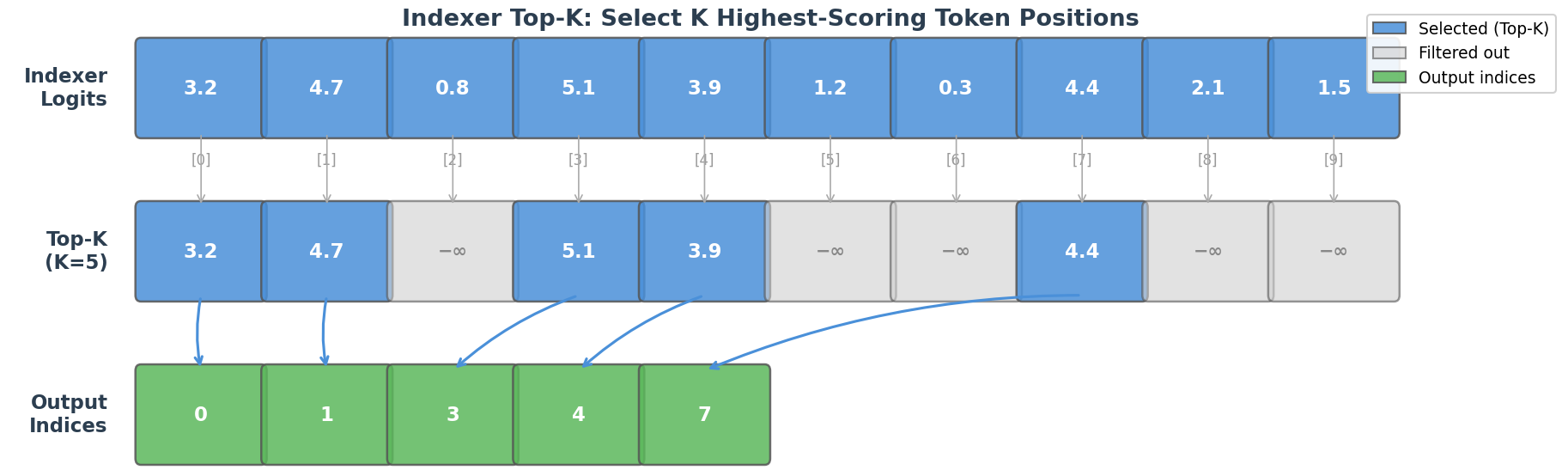}
  \caption{Indexer Top-K selection (illustrated with $K{=}5$, $N{=}10$). The indexer
    produces a score for each of the $N$ past tokens. The Top-K operator selects the $K$
    positions with the highest scores and outputs their indices. In DeepSeek Sparse
    Attention, $K{=}2048$ and $N$ can range from 8K to 128K+.}
  \label{fig:indexer-topk}
\end{figure}

The Top-K step is critical to the DSA pipeline: it must be both fast (to avoid becoming the
latency bottleneck) and correct (to preserve model accuracy). The challenge intensifies with
longer sequences---as $N$ grows from 8K to 64K or beyond, the Top-K kernel must process
proportionally more data while the output size ($K = 2048$) remains fixed.

\subsection{Existing Radix-Select-Based Top-K Implementation}
\label{sec:radix-select}

The production TensorRT-LLM radix-select baseline is an evolution of the parallel Top-K
framework presented by \citet{zhangsc23}, ported and further optimized for NVIDIA Blackwell
(sm\_100) by the same team. It represents the current state-of-the-art
\emph{distribution-agnostic} GPU Top-K implementation on this architecture, already
achieving 7.4$\times$ speedup over \texttt{torch.topk}. Note that alternative GPU Top-K
implementations such as RadiK~\citep{radik} do not have publicly available Blackwell
kernels, precluding direct comparison on the same platform.

The implementation (\texttt{invokeIndexerTopKDecode} in \texttt{indexerTopK.cu}) dispatches
to different kernel variants based on sequence length:

\begin{figure}[t]
  \centering
  \includegraphics[width=0.79\textwidth]{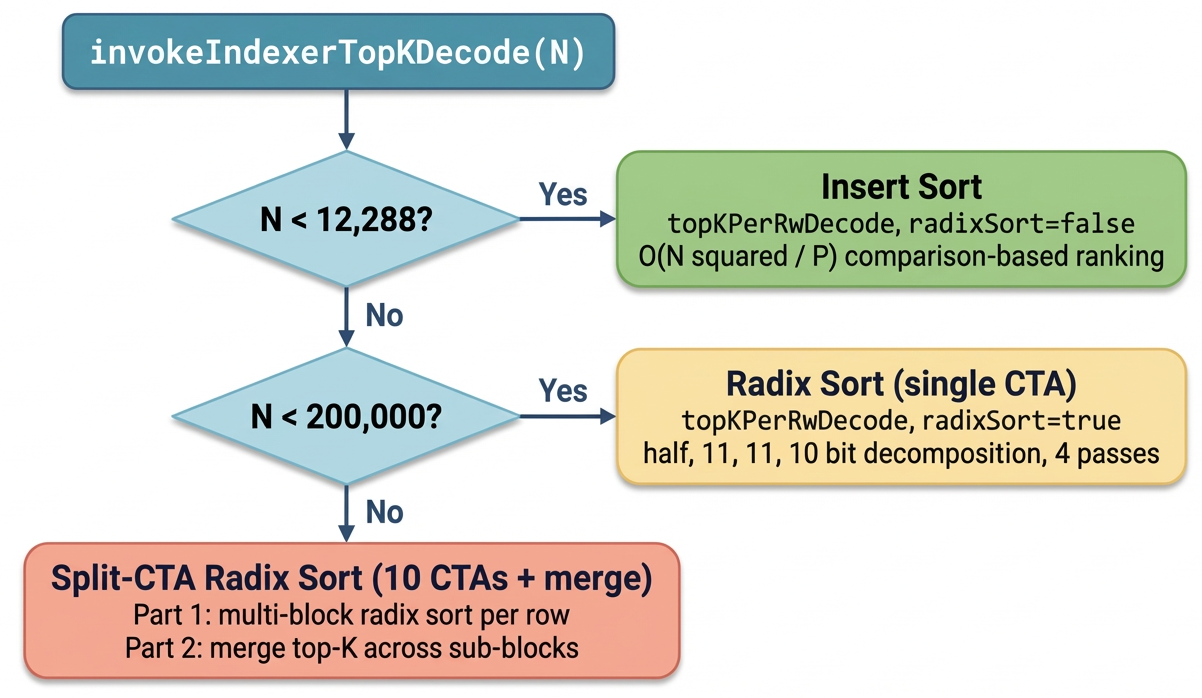}
  \caption{Original decode-stage Top-K dispatch in \texttt{invokeIndexerTopKDecode} (before
    GVR Top-K). The kernel selects insert sort for short sequences, radix sort for
    medium sequences, and a multi-CTA split approach for very long sequences.}
  \label{fig:dispatch-flowchart}
\end{figure}

The core algorithm (\texttt{topKPerRowJob}) uses a \textbf{radix-select} approach that is
data-distribution-agnostic. It partitions the 32-bit floating-point representation into
digit groups and iteratively narrows the candidate set:

\begin{enumerate}
  \item \textbf{Histogram}: Count elements per digit bucket using shared-memory atomic
    operations.
  \item \textbf{Prefix Sum}: Compute cumulative counts via \texttt{cub::BlockScan}.
  \item \textbf{Find Threshold}: Identify which bucket contains the $K$-th element.
  \item \textbf{Filter}: Retain candidates in the threshold bucket; emit elements in higher
    buckets directly.
\end{enumerate}

The implementation follows a \textbf{half $\to$ 11 $\to$ 11 $\to$ 10} bit decomposition (4
iterations), with an optimization that exits early and switches to CUB radix sort or
comparison-based ranking when the candidate set drops below 2048 elements. For very long
sequences ($N > 200$K), the split-CTA path distributes work across 10 CTAs and merges
results.

Here, \textbf{half} denotes a coarse 16-bit radix pass over the sortable FP32 score key,
followed by 11-, 11-, and 10-bit passes over the remaining key bits if early exit does not
trigger. In other words, the kernel always ranks FP32 indexer scores; this shorthand
describes only the radix bucket schedule, not a mixed-precision score representation or a
different precision for the Top-K count $K$.

\subsection{Complexity of Classical Top-K Approaches}
\label{sec:classical-complexity}

\begin{table}[H]
  \centering
  \caption{Complexity comparison of Top-K algorithms. $N$: sequence length, $K{=}2048$,
    $P$: thread count, $R$ or $I$: number of iterative passes. The last two columns separate
    GPU implementation fit from sensitivity to data distribution or prediction quality.}
  \label{tab:complexity-classical}
  \small
  \setlength{\tabcolsep}{4pt}
  \begin{tabular}{@{}lcccc@{}}
    \toprule
    \textbf{Algorithm} & \textbf{Time} & \shortstack[c]{\textbf{Global}\\\textbf{Passes}} & \shortstack[c]{\textbf{GPU}\\\textbf{Fit}} & \shortstack[c]{\textbf{Data}\\\textbf{Sensitivity}} \\
    \midrule
    \texttt{torch.topk} (sort) & $O(N \log N)$ & Multiple & General-purpose & Low \\
    Radix Select (TRT-LLM) & $O(R \cdot N/P)$ & $R \leq 4$ & Strong SIMD & Low \\
    Heap / Priority Queue & $O(N \log K)$ & 1 pass & Weak SIMD & Low \\
    \textbf{GVR Top-K (ours)} & $O((I{+}1) \cdot N/P)$ & $I{+}1$ & Strong SIMD & High \\
    \bottomrule
  \end{tabular}
\end{table}

Table~\ref{tab:complexity-classical} summarizes the complexity of classical Top-K
approaches alongside GVR. The key distinction of the heuristic approach is that the
effective number of global-memory passes $I$ depends on the quality of the initial threshold
estimate: unlike radix-select, GVR preserves strong GPU parallelism but benefits most when
the previous-step prediction is informative, which, as we will show, is consistently true in
LLM decoding workloads.

\subsection{The Long-Sequence Bottleneck}
\label{sec:long-sequence}

Modern agentic AI systems routinely process contexts of 32K--128K tokens or more:
\begin{itemize}
  \item \textbf{Multi-turn tool use}: An agent accumulating conversation history, tool
    outputs, and intermediate reasoning across dozens of interaction rounds.
  \item \textbf{Long-document reasoning}: Summarization, QA, and analysis over extensive
    codebases, legal documents, or scientific papers.
  \item \textbf{Code generation}: Large repository contexts with cross-file dependencies.
\end{itemize}
These workloads stress every component of the inference pipeline that scales with $N$. While
the sparse MLA kernel benefits from token sparsity (computing attention only over $K =
2048$ tokens regardless of $N$), and the indexer MQA kernel has been optimized with FP8
arithmetic and Blackwell-specific instructions, the Top-K selector must still scan all $N$
indexer scores every decoding step.

The three DSA decode-step components have fundamentally different scaling characteristics.
Since all three are memory-bound during decode, the roofline cost of each is proportional
to its total memory traffic:

\begin{table}[t]
  \centering
  \caption{Scaling characteristics of DSA decode-step components. $N$: sequence length,
    $K{=}2048$, $R \approx 3$: radix-select passes, $d_i{=}128$: indexer head dimension,
    $d{=}192$: MLA head dimension (128 non-PE $+$ 64 PE).}
  \label{tab:scaling-components}
  \begin{tabular}{@{}lccc@{}}
    \toprule
    \textbf{Component} & \textbf{Scaling} & \textbf{Total Memory Traffic} & \textbf{Trend as $N$ Grows} \\
    \midrule
    Indexer MQA    & $O(N)$       & $N \cdot d_i \cdot 2\text{B}$    & Linear growth \\
    Top-K (radix)  & $O(R \cdot N)$ & $R \cdot N \cdot 4\text{B}$     & Linear ($R$ passes) \\
    Sparse MLA     & $O(K)$       & $K \cdot d \cdot 2\text{B}$      & \textbf{Constant} ($K$ fixed) \\
    \bottomrule
  \end{tabular}
\end{table}

The Top-K memory traffic $R \cdot N \cdot 4\text{B}$ grows linearly with $N$ while sparse
MLA remains constant at $K \cdot d \cdot 2\text{B}$. This means the \textbf{Top-K fraction
of DSA latency increases monotonically} with sequence length---from a minor component at
short sequences to a major decode-time bottleneck at long sequences. The production
radix-select kernel already achieves 7.4$\times$ over \texttt{torch.topk}, yet two dominant
contributors still limit its efficiency in our measurements beyond the raw $O(R \cdot N)$
traffic:

\begin{itemize}
  \item \textbf{Multi-pass data re-reads}: Each of the $R \approx 3$ radix-select steps
    performs two full scans of all $N$ elements (histogram build + filter/collect), totaling
    ${\sim}6$ $N$-element scans per kernel invocation.
  \item \textbf{Shared-memory atomic serialization}: The histogram phase uses
    \texttt{atomicAdd} on shared-memory bin counters (2048 bins, hot-bin contention), and
    the collect phase funnels all qualifying elements through a single
    \texttt{atomicAdd(\&counter, 1)}---serializing threads that compete on the same address
    and reducing effective SIMT utilization well below the memory bandwidth ceiling.
\end{itemize}

The scan counts above refer to \emph{logical} full-row traversals. Under cold-start timing,
the first traversal comes from DRAM, while later re-scans may partially benefit from L2
reuse; however, each pass still reprocesses the full $N$-element score row and therefore
remains a first-order contributor to both memory traffic and latency.

This motivates the search for a Top-K algorithm that can reduce the number of
global-memory passes by exploiting properties specific to LLM decoding.

\section{Temporal Correlation in LLM Sparse Attention}
\label{sec:temporal-correlation}

\subsection{Empirical Observation}
\label{sec:empirical-observation}

A key empirical observation is that the Top-K index sets exhibit \textbf{strong temporal
correlation} during LLM decoding: the set of important tokens at step $t$ overlaps
significantly with the set at step $t-1$. We measured this overlap (hit ratio) on real
DeepSeek-V3.2 decode-stage indexer logits from \textbf{SWE-bench-derived LongSeqTasks}, a
public long-context prompt collection constructed from handpicked SWE-bench
tasks~\citep{swebench} selected for long ISL or long OSL. We release these
prompt buckets together with the main-result reproduction scripts in a public
supplementary artifact repository:
\href{https://github.com/longcheng-nv/GVR_TopK_supplementaty_materials}{supplementary GitHub repository}.
The two public prompt buckets themselves,
\texttt{swe\_bench\_64k.jsonl} and \texttt{swe\_bench\_100k.jsonl}, are available in the
\href{https://github.com/longcheng-nv/GVR_TopK_supplementaty_materials/tree/main/longseqtasks}{LongSeqTasks subdirectory}.

\begin{figure}[t]
  \centering
  \includegraphics[width=0.92\textwidth]{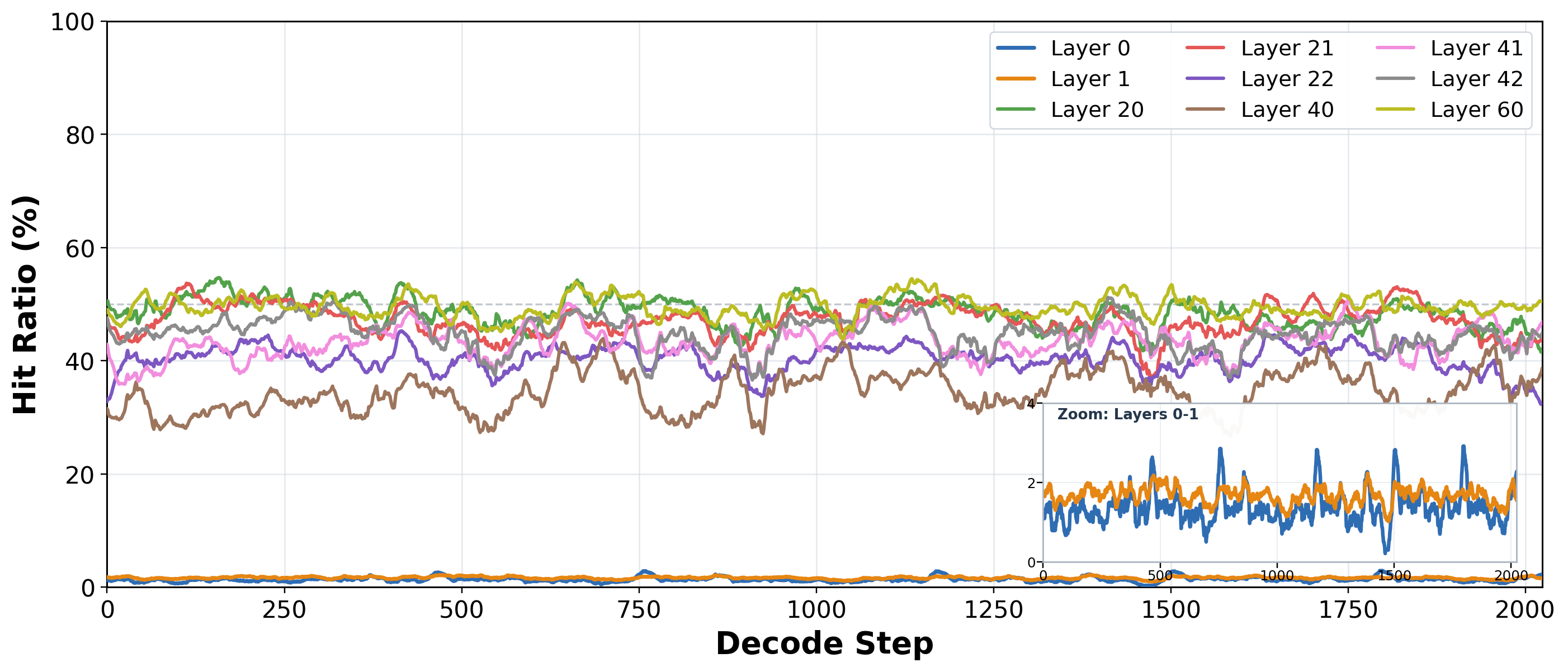}
  \caption{Raw Top-K overlap (hit ratio) between consecutive decoding steps across different
    layers in DeepSeek-V3.2, measured on entry 1 of
    \texttt{swe\_bench\_64k.jsonl} from SWE-bench-derived LongSeqTasks (2,025 generated
    tokens; 2,024 valid previous-step comparisons after skipping step 0).
    Here ``raw'' means that the current-step Top-K index set is compared directly against the
    previous-step Top-K index set without any positional shift. For readability, each curve is
    rendered with a 25-step moving average over decode steps, and the inset zooms into the
    0--4\% range for Layers 0 and 1. Layers 20--60
    exhibit 35--50\% average overlap (max ${\sim}$60\%), while Layers 0 and 1 show
    near-zero overlap (${\sim}$1--2\%), reflecting distinct indexer score dynamics in early
    vs.\ deeper layers.}
  \label{fig:hit-ratio}
\end{figure}

Figure~\ref{fig:hit-ratio} reports \emph{raw} overlap, which is the strictest notion of
temporal persistence. A complementary view is \emph{shifted} overlap: when the query advances
by one step, the previous-step indices are shifted by $+1$ before comparison. The shifted
view is not used by the current GVR predictor directly, but it is helpful for visualizing the
Toeplitz-induced translation of the score landscape and for explaining why a nearby Top-K set
remains a useful warm-start even when the raw overlap of exact indices is modest in some
layers.

\begin{figure}[t]
  \centering
  \includegraphics[width=0.90\textwidth]{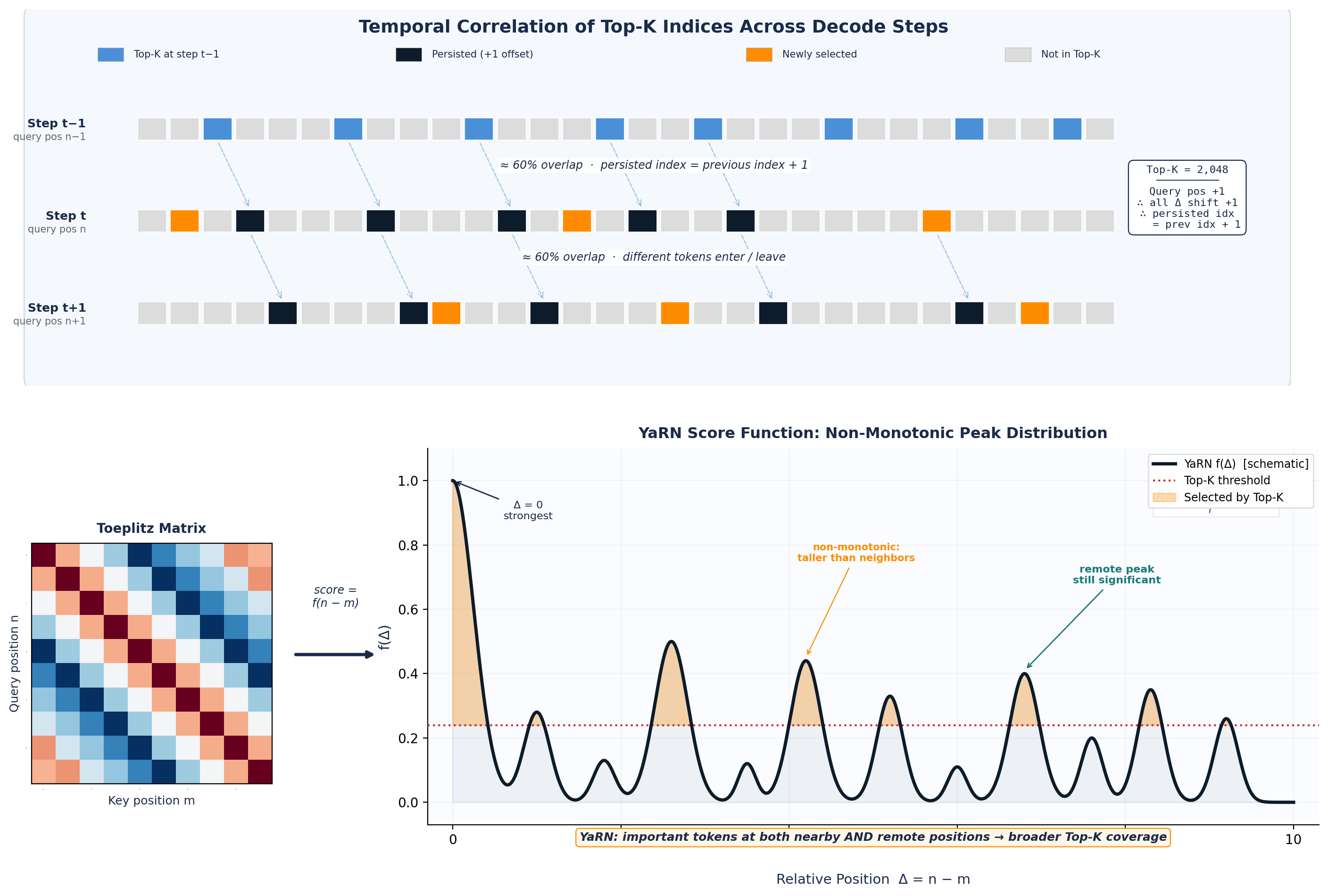}
  \caption{Top: Temporal correlation under an offset+1 shift---when the query advances by one
    position, all relative positions $\Delta$ shift by $+1$, causing ${\sim}$60\% of Top-K
    tokens to persist (dark navy, shifted right by one cell) while ${\sim}$40\% change
    (orange). Bottom-left: Toeplitz structure of the attention score matrix. Bottom-right:
    Normalized $g(\Delta)$ comparison---standard RoPE (blue dashed) shows rapid peak decay,
    while YaRN (dark solid) maintains significant peaks at large $\Delta$, enabling Top-K
    selection across both nearby and remote positions.}
  \label{fig:temporal-correlation}
\end{figure}

This temporal correlation is not coincidental---it has a principled theoretical basis in the
structure of the indexer's attention mechanism.

\subsection{Theoretical Basis: Toeplitz Structure and RoPE Frequency Analysis}
\label{sec:rope-theory}

The DSA indexer computes token importance scores via an MQA dot product between
RoPE-encoded~\citep{rope} query and key tensors. In RoPE, each query-key pair at positions
$(n, m)$ is multiplied by a block-diagonal rotation matrix $R_{n-m}$ whose entries are
cosines and sines of position-dependent angles. For the indexer's RoPE-only dimensions
($d_{\mathrm{rope}} = 64$ in DeepSeek-V3.2), the positional contribution to the attention
score reduces to:
\begin{equation}
  g(\Delta) = 2 \sum_{i=0}^{d_{\mathrm{rope}}/2 - 1} \cos(\Delta \cdot \theta_i),
  \quad \Delta = n - m,
  \quad \theta_i = \beta^{-2i/d_{\mathrm{rope}}}
  \label{eq:g-delta}
\end{equation}
where $\beta = 10000$ is the RoPE base frequency. This expression is the inner product of
all-ones vectors transformed by $R_\Delta$, representing the pure positional-encoding
contribution independent of data content.

\paragraph{Toeplitz property.}
Since $g$ depends only on the relative position $\Delta = n - m$, the positional score
matrix $P \in \mathbb{R}^{S \times S}$ is a Toeplitz matrix---constant along each diagonal.
This reduces the problem of understanding which token pairs are positionally favored from a
2D matrix analysis to a \textbf{1D function analysis} on $g(\Delta)$.

\paragraph{Multi-scale cosine superposition.}
$g(\Delta)$ is a superposition of $d_{\mathrm{rope}}/2 = 32$ cosines with periods spanning
$2\pi/\theta_0 \approx 6.3$ to $2\pi/\theta_{31} \approx 58{,}600$---a 10{,}000:1
frequency ratio. The global maximum is at $\Delta = 0$ (self-position); secondary peaks
arise at positions where cosine waves constructively interfere. Because $g$ is smooth,
advancing the query by one position ($\Delta \to \Delta + 1$) causes only a small
perturbation to the score landscape---providing a structural rationale for why Top-K
indices often change slowly between consecutive decode steps.

\paragraph{YaRN extension.}
DeepSeek-V3.2 extends RoPE with YaRN~\citep{yarn} (scaling factor $s = 40$), which
interpolates low-frequency components while preserving high-frequency ones. The effect on
$g(\Delta)$ is that \textbf{peaks at large relative positions remain significant} rather
than decaying monotonically (see Figure~\ref{fig:temporal-correlation}, bottom-right). This
produces a more spatially uniform distribution of important tokens---favoring the heuristic
Top-K approach, since the Top-K set spans both nearby and distant positions, providing a
richer prediction signal for the next step.

\subsection{Pre-Computed Candidate Indices}
\label{sec:precomputed-candidates}

The frequency structure of RoPE enables a static pre-computation. Given the model's RoPE
parameters (dimension $d = 64$, YaRN scaling with \texttt{scaling\_factor\,=\,40}), we can
compute the idealized score function $g(\Delta)$ for all possible relative positions and
find the $K$ positions with the largest scores:
\begin{equation}
  \mathcal{P}_{\mathrm{static}} = \operatorname{argtopk}_{\Delta \geq 0} \; g(\Delta)
  \label{eq:static-preidx}
\end{equation}

We evaluated several candidate strategies---including peak indices of $g(\Delta)$ and direct
Top-K of $g(\Delta)$---and found that TopK-based prediction significantly outperforms
peak-index approaches (see Appendix~\ref{app:preidx-analysis} for the comparative analysis).

This pre-computed index set captures the \textbf{structural prior}---the positions that RoPE
frequency structure inherently favors. During inference, the actual Top-K indices for a
query at position $n$ are the positions $m$ such that $n - m \in \mathcal{P}_{\mathrm
{static}}$, modulated by the data-dependent content of the actual Q/K tensors and the
indexer weight~$W^I$.

In practice, we use the \textbf{previous step's Top-K result} as the prediction signal
(\texttt{preIdx}), which captures both the structural RoPE prior and the data-dependent
content correlation. For the majority of layers (L20--L60), this achieves prediction
accuracy $\alpha \approx 0.35$--$0.50$ (35--50\% of the previous Top-K indices remain in
the current Top-K set), which empirically is often enough for the heuristic algorithm to
converge in 1--2 interpolation iterations. Notably, Layers~0 and~1 exhibit near-zero temporal correlation
($\alpha \approx 0.01$), causing the heuristic kernel to fall back on more interpolation
iterations for those layers---consistent with their lower speedup ratios in the benchmarks.

\section{GVR: Guess-Verify-Refine Top-K Algorithm}
\label{sec:algorithm}

\subsection{Core Idea}
\label{sec:core-idea}

Given input vector $\mathbf{x} = (x_0, x_1, \ldots, x_{N-1}) \in \mathbb{R}^N$ and a
predicted index set $\mathcal{P} = \{p_0, p_1, \ldots, p_{M-1}\} \subset \{0, \ldots,
N{-}1\}$ (where $M = 2048$), find index set $\mathcal{S}^\ast$ with $|\mathcal{S}^\ast| =
K$ containing the indices of the $K$ largest values in~$\mathbf{x}$.

\paragraph{Candidate-set validity.}
Let $x_{(1)} \ge x_{(2)} \ge \cdots \ge x_{(N)}$ denote the order statistics of
$\mathbf{x}$ in descending order, and define the threshold-counting function
\[
  f(T) = |\{\, i \in \{0,\ldots,N-1\} \mid x_i \ge T \,\}|.
\]
The exactness argument below is conditional on finding a threshold whose candidate set is
both inclusive enough to contain the true Top-K and small enough to fit the refinement
buffer. Lemma~1 formalizes this conditional guarantee.

\smallskip
\noindent\textbf{Lemma 1 (Top-K containment).}
Let $\mathcal{S}^\ast$ be any valid exact Top-K solution of size $K$. If
\[
  K \le f(T) \le C,
\]
where $C = \texttt{MAX\_CANDIDATES} = 6144$, then
$\mathcal{S}^\ast \subseteq \{\, i \mid x_i \ge T \,\}$.

\smallskip
\noindent\textit{Proof.}
Since $f(T) \ge K$, the threshold satisfies $T \le x_{(K)}$.
Therefore every element in any exact Top-K solution has value at least
$x_{(K)} \ge T$, and must belong to $\{\, i \mid x_i \ge T \,\}$. The upper bound
$f(T) \le C$ ensures that the candidate set fits within the
refinement capacity. \hfill$\square$

Thus, once a threshold satisfying $K \le f(T) \le C$ has been found, the subsequent
Verify/Refine stages are exact. Lemma~1 by itself does not guarantee that every input admits
such a threshold for a fixed refinement capacity~$C$; rather, it isolates the condition under
which the downstream exact-selection argument holds.

The algorithm uses the predicted indices $\mathcal{P}$ to estimate a threshold $T$ that is
close to the true $K$-th largest value $x_{(K)}$. With a good estimate, only 1--2 global
passes over the data are needed (versus 3--4 in radix select), followed by
in-shared-memory refinement on the small candidate set.

This four-phase pipeline gives rise to the \textbf{Guess-Verify-Refine (GVR)}
algorithm. \textbf{Guess} (Phases~1--2) uses the previous step's Top-K to
estimate a tight threshold via secant interpolation. \textbf{Verify}
(Phase~3) collects all candidates above the threshold into shared memory,
confirming that the candidate set contains the true Top-K. \textbf{Refine}
(Phase~4) performs exact selection within the small candidate set using
histogram-based in-shared-memory operations.

\subsection{Algorithm Phases}
\label{sec:phases}

\begin{figure}[t]
  \centering
  \includegraphics[width=0.90\textwidth]{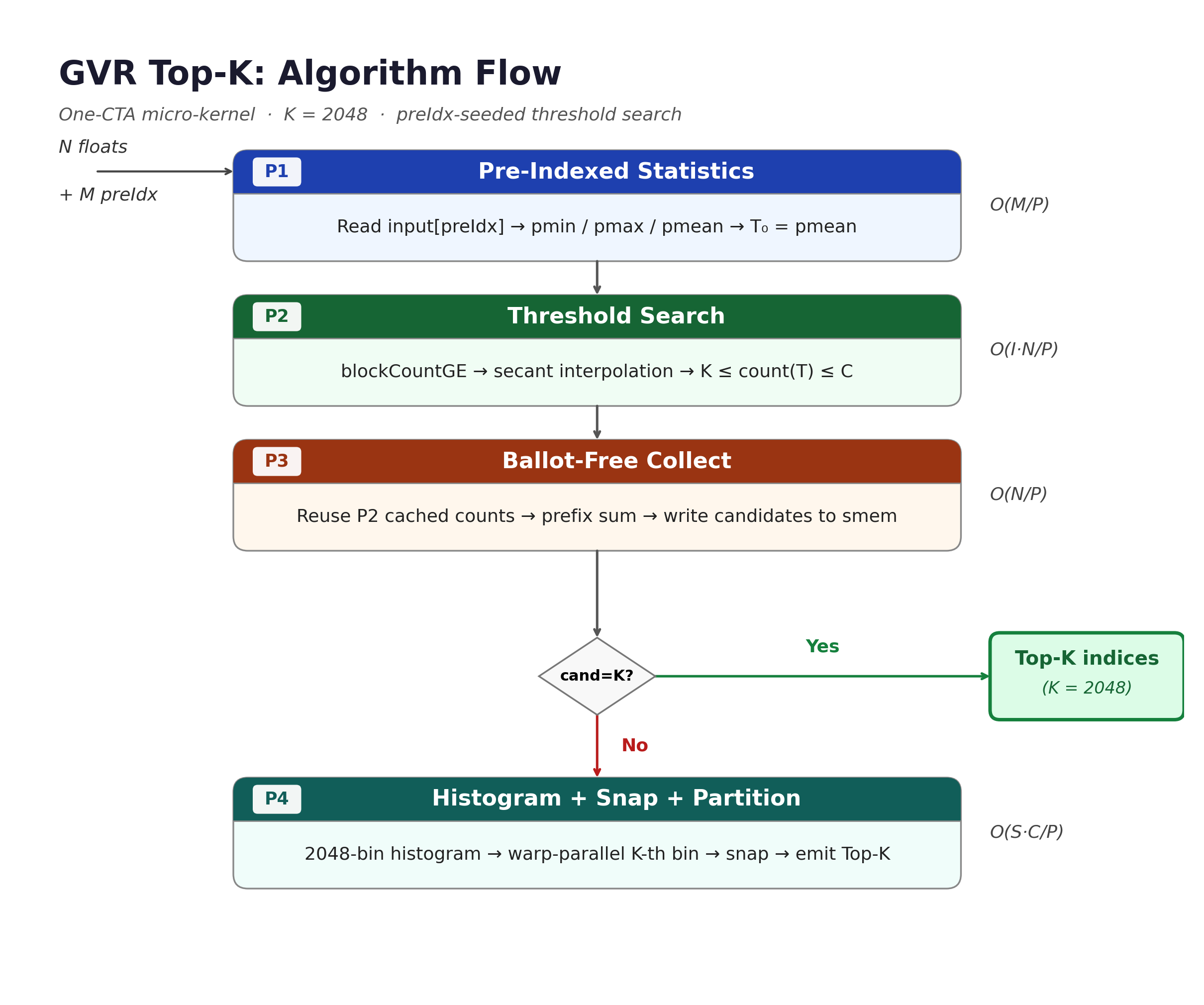}
  \caption{GVR Top-K algorithm flow: \textbf{Guess} (Phases~1--2) estimates a threshold from
    the previous step's Top-K, \textbf{Verify} (Phase~3) collects candidates above the
    threshold, \textbf{Refine} (Phase~4) selects exactly $K$ elements. All four phases
    execute sequentially within a single CTA. Complexity annotations show per-phase cost;
    $I$ and $S$ denote the number of threshold-search and snap iterations, respectively.}
  \label{fig:algorithm-flow}
\end{figure}

\subsubsection{Phase 1: Pre-Indexed Statistics}
\label{sec:phase1}

Using the predicted index set $\mathcal{P}$, compute the min, max, and mean of the
corresponding input values:
\begin{equation}
  T_0 = \bar{x}_{\mathcal{P}} = \frac{1}{|\mathcal{P}|} \sum_{i \in \mathcal{P}} x_i
  \label{eq:t0}
\end{equation}

Let
\[
  \alpha = \frac{|\mathcal{P} \cap \mathcal{S}^\ast|}{|\mathcal{P}|},
\]
denote the hit ratio of the prediction set. In our implementation,
$|\mathcal{P}| = M = K = 2048$, so this is also the fraction of previous-step Top-K
indices that remain in the current exact Top-K. Define the empirical means of the hit and
miss subsets as
\[
  \mu_{\mathrm{hit}}
  = \frac{1}{|\mathcal{P} \cap \mathcal{S}^\ast|}
      \sum_{i \in \mathcal{P} \cap \mathcal{S}^\ast} x_i,
  \qquad
  \mu_{\mathrm{miss}}
  = \frac{1}{|\mathcal{P} \setminus \mathcal{S}^\ast|}
      \sum_{i \in \mathcal{P} \setminus \mathcal{S}^\ast} x_i .
\]
Partitioning $\mathcal{P}$ into hits and misses then gives the exact identity
\begin{equation}
  T_0
  = \alpha \cdot \mu_{\mathrm{hit}} + (1 - \alpha) \cdot \mu_{\mathrm{miss}} .
  \label{eq:t0-decomp}
\end{equation}
Next define the empirical means of the true Top-K and non-Top-K populations:
\[
  \mu_{\mathrm{top}}
  = \frac{1}{K} \sum_{i \in \mathcal{S}^\ast} x_i,
  \qquad
  \mu_{\mathrm{non}}
  = \frac{1}{N-K} \sum_{i \notin \mathcal{S}^\ast} x_i .
\]
Under the representativeness approximation that the hit subset
$\mathcal{P} \cap \mathcal{S}^\ast$ is representative of the true Top-K population and the
miss subset $\mathcal{P} \setminus \mathcal{S}^\ast$ is representative of the non-Top-K
population, we have $\mu_{\mathrm{hit}} \approx \mu_{\mathrm{top}}$ and
$\mu_{\mathrm{miss}} \approx \mu_{\mathrm{non}}$, yielding
\begin{equation}
  T_0 \approx \alpha \cdot \mu_{\mathrm{top}} + (1 - \alpha) \cdot \mu_{\mathrm{non}} .
  \label{eq:t0-approx}
\end{equation}
\noindent
The unconditional mean of the full input satisfies
\[
  \mu_{\mathrm{all}}
  = \frac{1}{N}\sum_{i=0}^{N-1} x_i
  = \frac{K}{N} \cdot \mu_{\mathrm{top}}
    + \left(1 - \frac{K}{N}\right) \cdot \mu_{\mathrm{non}} .
\]
Subtracting $\mu_{\mathrm{all}}$ from~\eqref{eq:t0-approx} gives
\begin{equation}
  T_0 - \mu_{\mathrm{all}}
  \approx \left(\alpha - \frac{K}{N}\right)
    \left(\mu_{\mathrm{top}} - \mu_{\mathrm{non}}\right) .
  \label{eq:t0-gap}
\end{equation}
For any non-degenerate Top-K instance, $\mu_{\mathrm{top}} > \mu_{\mathrm{non}}$.
Therefore, whenever $\alpha > K/N$, the pre-index mean $T_0$ is expected to exceed the
unconditional mean. In our decode regime, $K = 2048$ while $N \ge 8192$, so $K/N \le 0.25$
and decreases with sequence length; at the same time, most layers exhibit observed hit
ratios $\alpha \approx 0.35$--$0.50$. Thus $T_0$ is typically above the unconditional mean,
which helps move the initial threshold toward the Top-K tail containing $x_{(K)}$ and
empirically supports faster Phase~2 convergence. The phase uses scattered \texttt{\_\_ldg} reads and
\texttt{redux.sync} warp reductions (a single instruction on sm\_80+ replacing 5 shuffle
operations).

\textbf{Cost}: $O(M/P)$---reading $M = 2048$ scattered global memory values with $P = 512$
threads.

\subsubsection{Phase 2: Secant-Method Threshold Search}
\label{sec:phase2}

\begin{figure}[t]
  \centering
  \includegraphics[width=0.88\textwidth]{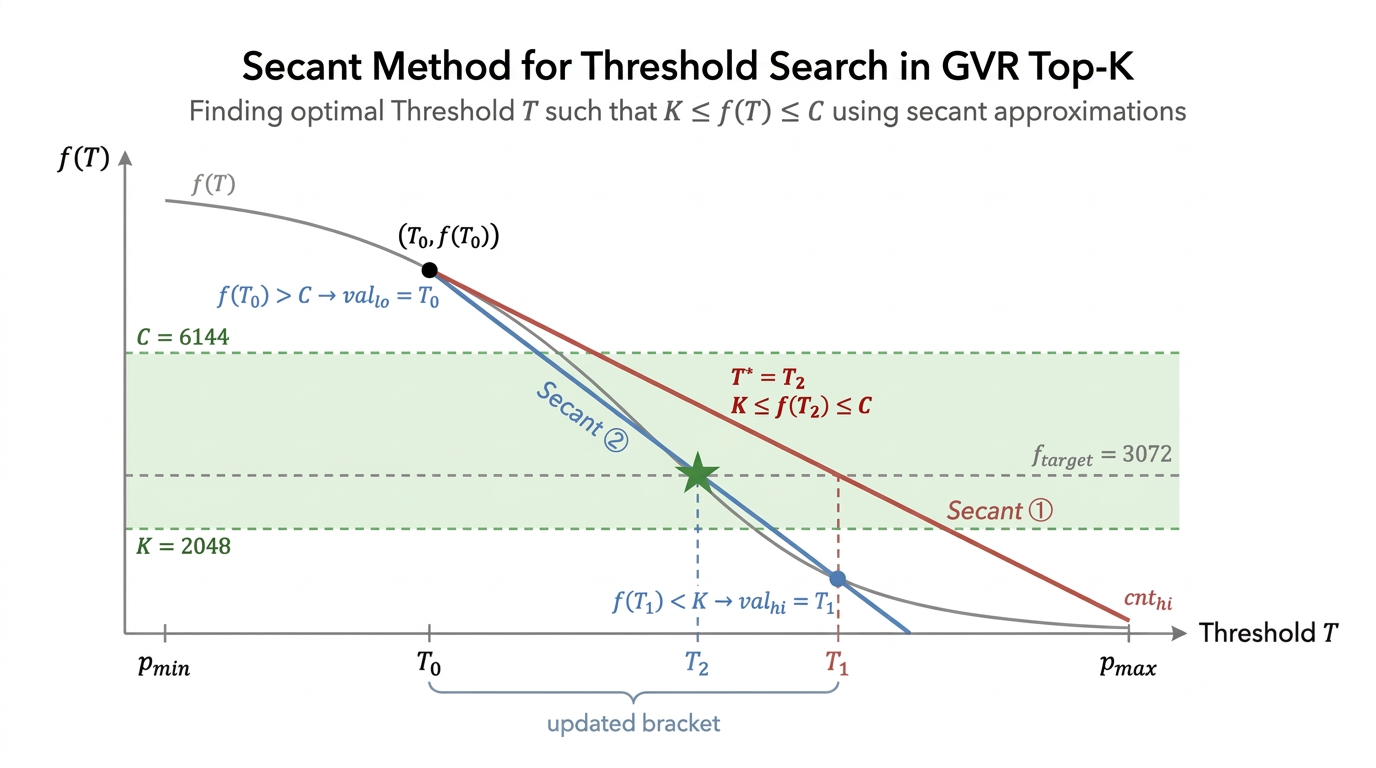}
  \caption{Phase~2 interpolation-based threshold search. Starting from
    $T_0 = \texttt{pmean}$ with bracket $[\texttt{pmin}, \texttt{pmax}]$, the algorithm
    evaluates $f(T_0)$: since $f(T_0) > C$, \texttt{val\_lo} is set to $T_0$. Secant~1
    connects $(\texttt{val\_lo}, \texttt{cnt\_lo})$ and $(\texttt{val\_hi},
    \texttt{cnt\_hi})$, crossing $f_{\mathrm{target}}$ to determine $T_1$. Since
    $f(T_1) < K$, \texttt{val\_hi} is updated to $T_1$, narrowing the bracket. Secant~2
    connects the updated anchors and crosses $f_{\mathrm{target}}$ to produce $T_2$, which
    lands in the target zone $[K, C]$ in this illustrative example. First-iteration damping
    ($f \leq 0.50$) prevents overshoot.}
  \label{fig:secant-method}
\end{figure}

The threshold-counting function $f(T) = |\{i : x_i \geq T\}|$ is a monotonically
non-increasing step function. The goal is to find $T^\ast$ such that
$K \leq f(T^\ast) \leq C$.

Each iteration computes an exact global count via
\texttt{block\-Count\-GE}---a bandwidth-bound loop using \texttt{float4}
vectorized \texttt{\_\_ldg} loads, pure register comparison and accumulation,
and warp-level reduction. Critically, \texttt{block\-Count\-GE} caches each
thread's partial count in a per-thread shared-memory slot
(\texttt{per\_thread\_counts[tid]}) before the warp reduction. This cache is
reused by Phase~3 to eliminate a redundant $N$-scan (see below).

The threshold update uses secant interpolation:
\begin{equation}
  T_{\mathrm{new}} = T_{\mathrm{lo}}
    + \frac{f(T_{\mathrm{lo}}) - f_{\mathrm{target}}}
           {f(T_{\mathrm{lo}}) - f(T_{\mathrm{hi}})}
    \cdot (T_{\mathrm{hi}} - T_{\mathrm{lo}})
  \label{eq:secant}
\end{equation}
with first-iteration damping ($\leq 0.5$) to prevent overshoot, and bisection fallback at
float precision limits. Because $f$ is a discrete step function, classical secant-method
convergence theory does not directly apply here. We instead use secant interpolation as a
heuristic inverse-CDF estimate near the target region, and treat the observed 1--2
iteration behavior on real decode workloads as an empirical property rather than a proved
rate guarantee.

When Phase~2 converges cleanly (\texttt{done=1}), a \textbf{safety-net guard} skips the
subsequent verification \texttt{blockCountGE} call that would otherwise be redundant, saving
${\sim}$4\,$\mu$s per kernel invocation.

\textbf{Cost}: $O(I \cdot N/P)$ where $I$ is the number of secant iterations. A
kernel-faithful replay of the Phase~2 control flow over 18{,}207 real decode
invocations from entry~1 of the 64K bucket in SWE-bench-derived LongSeqTasks
(9 layers $\times$ 2{,}023 valid decode steps, using the previous step's
\texttt{preIdx}) shows that Phase~2 finishes in a single
iteration 67.6\% of the time, within 2/3/4 iterations 84.3\%/94.8\%/99.4\% of
the time, and never triggers the \texttt{done=2} fallback path; the maximum
observed is 6 iterations. High-correlation layers (L20--60) average 1.1--1.6
iterations per invocation, whereas low-correlation L0--1 average 2.1--2.7. For
comparison, synthetic data with poor prediction typically requires 4--6
iterations.

\subsubsection{Phase 3: Ballot-Free Candidate Collection}
\label{sec:phase3}

Once a valid threshold is found, all elements $\geq T$ are collected into shared memory.
This phase uses a \textbf{ballot-free} design with a key optimization:
\textbf{Phase~3 sub-pass~1 (count) is eliminated} by reusing the per-thread
counts cached by Phase~2's last \texttt{blockCountGE} call. Since the threshold
has not changed between Phase~2's final count and Phase~3, the cached
\texttt{per\_thread\_counts[tid]} values remain valid, saving an entire
$N$-element rescan (${\sim}$4\,$\mu$s).

\begin{itemize}
  \item \textbf{Write-offset computation}: Read per-thread counts from cache $\to$ warp
    prefix sum $\to$ cross-warp prefix sum $\to$ pre-computed write offsets.
  \item \textbf{Sub-pass~2 (Write)}: Each thread writes to its pre-allocated shared-memory
    range. No \texttt{\_\_ballot\_sync}, no \texttt{\_\_shfl\_sync}, no \texttt{atomicAdd}.
\end{itemize}

The ballot-free design is critical: \texttt{\_\_ballot\_sync} acts as a compiler barrier
that serializes L2 load pipelining. The original per-element ballot approach costs
${\sim}$30{,}000 cycles (5$\times$ the \texttt{blockCountGE} cost); the ballot-free
approach costs ${\sim}$16{,}000 cycles.

\textbf{Cost}: $O(N/P)$---one sub-pass scanning the full input (the count sub-pass is
eliminated via caching).

\subsubsection{Phase 4: Histogram-Based Exact Selection}
\label{sec:phase4}

\begin{figure}[t]
  \centering
  \includegraphics[width=0.90\textwidth]{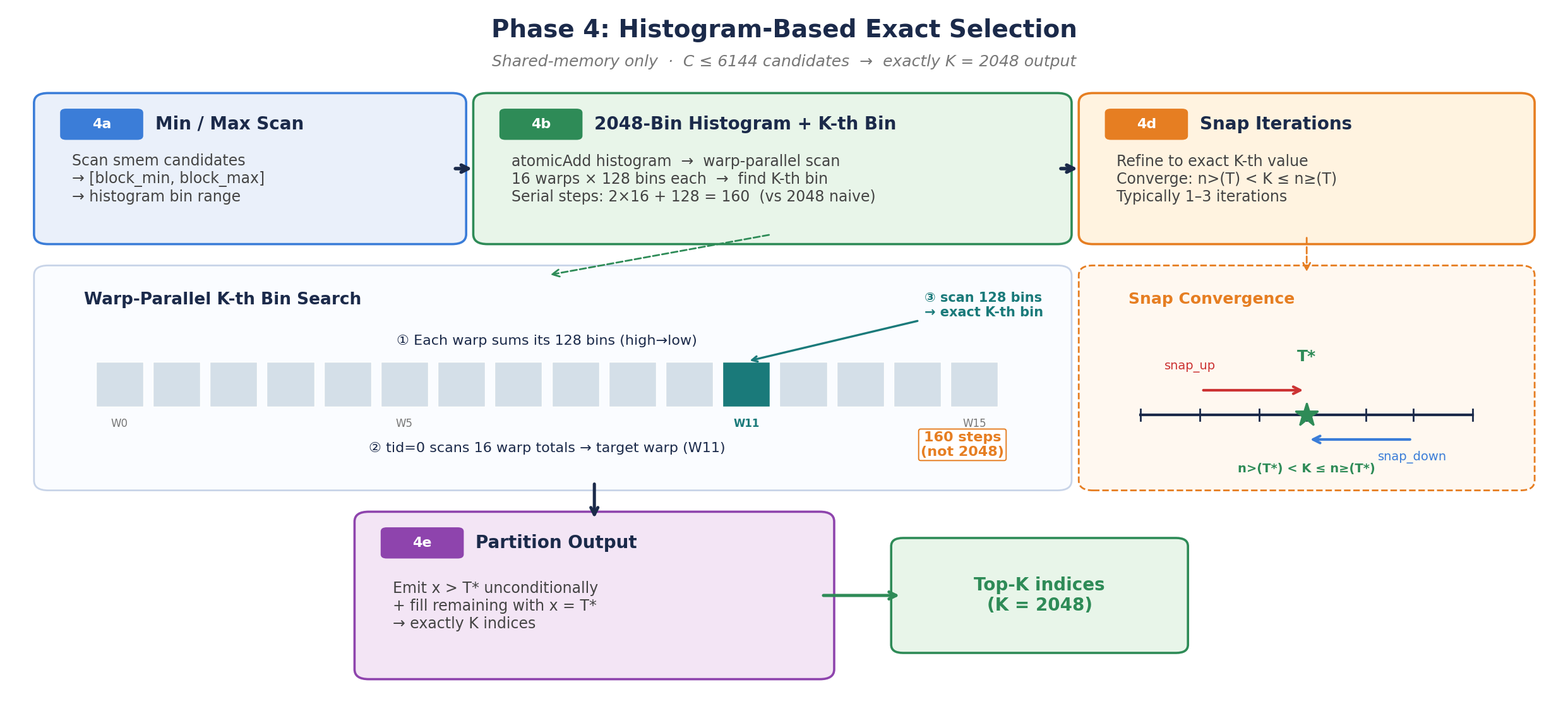}
  \caption{Phase~4 detail: 4a scans candidates for min/max range; 4b builds a 2048-bin
    histogram with warp-parallel $K$-th bin search (16 warps $\times$ 128 bins, 160 serial
    steps); 4d refines via snap iterations converging when $n_{>}(T) < K \leq n_{\geq}(T)$;
    4e partitions the final output.}
  \label{fig:phase4}
\end{figure}

If the candidate count does not exactly equal $K$, a shared-memory refinement selects
exactly $K$ elements:

\begin{enumerate}
  \item \textbf{Min/Max scan} over candidates for accurate histogram bins.
  \item \textbf{2048-bin histogram} via \texttt{atomicAdd} over the candidate set, followed
    by a \textbf{warp-parallel $K$-th bin search}: each warp sums its
    $\texttt{NUM\_BINS}/\texttt{NUM\_WARPS}$ bins (high-to-low), then \texttt{tid=0} scans
    \texttt{NUM\_WARPS} warp-totals to find the target warp, and finally the target warp's
    \texttt{lane=0} locates the exact bin. This reduces serial scan steps from 2048 to
    $2 \times \texttt{NUM\_WARPS} + \texttt{NUM\_BINS}/\texttt{NUM\_WARPS} = 160$
    (12.8$\times$ fewer).
  \item \textbf{Snap iterations}: Refine the threshold to the exact $K$-th largest
    value by stepping through distinct data values. Each fused snap iteration computes
    (\texttt{count\_ge}, \texttt{count\_gt}, \texttt{snap\_up}, \texttt{snap\_down})
    in one shared-memory scan. Convergence: when $n_{>}(T) < K \leq n_{\geq}(T)$.
    On the same 18{,}207-invocation replay used for Phase~2, and conditioned on
    candidate sets with \texttt{cand\_count} $> K$, snap still converges quickly:
    79.3\% of cases finish within 5 iterations and 95.3\% within 8, with a
    maximum observed tail of 20.
  \item \textbf{Partition}: Emit elements $> T^\ast$ unconditionally, fill remaining slots
    with elements $= T^\ast$.
\end{enumerate}

\textbf{Cost}: $O(S \cdot C/P)$ where $S$ is the number of snap iterations and
$C \leq 6144$ candidates. The same replay shows layer-wise averages of
2.2--4.7 snap iterations (lowest on L1, highest on L60). This is purely
shared-memory work---no global memory access. Because each snap iteration scans
only the candidate buffer (typically ${\sim}$2.6K--3.8K elements) rather than
the full $N{\sim}70$K input, its data movement is still roughly 23$\times$
lower than an additional full-row count pass.

\subsection{Complexity Analysis}
\label{sec:complexity-analysis}

\begin{table}[t]
  \centering
  \caption{Per-phase complexity of the GVR Top-K algorithm. $M{=}2048$: prediction set
    size, $N$: sequence length, $P{=}512$: thread count, $I$: threshold-search iterations,
    $S$: snap iterations, $C \leq 6144$: candidate count, $N_{\mathrm{bin}}{=}2048$:
    histogram bins.}
  \label{tab:phase-complexity}
  \small
  \begin{tabular}{@{}lccl@{}}
    \toprule
    \textbf{Phase} & \textbf{Time Complexity} & \textbf{Memory Access} & \textbf{Space} \\
    \midrule
    P1 (PreIdx Stats)       & $O(M/P)$             & $M$ scattered global        & $O(1)$ regs \\
    P2 (Threshold Search)   & $O(I \cdot N/P)$     & $I \!\times\! N$ seq.\ (L2) & $O(1)$ regs \\
    P3 (Candidate Collect)  & $O(N/P)$             & $N$ sequential (L2)         & $O(C)$ smem \\
    P4 (Exact Selection)    & $O(S \cdot C/P)$     & Shared memory only          & $O(N_{\mathrm{bin}})$ smem \\
    \midrule
    \textbf{Total} & $O\!\bigl((I{+}1)\!\cdot\! N/P {+} S\!\cdot\! C/P\bigr)$ & $(I{+}1)\!\times\! N {+} M$ & ${\sim}$60\,KB \\
    \bottomrule
  \end{tabular}
\end{table}

For real decoding data ($I \approx 2$, $S \approx 2$, $P = 512$, $N_{\mathrm{bin}} = 2048$, $C \leq
6144$): approximately $3N/P + 2C/P$ memory accesses. The Phase~3 count-cache optimization
eliminates one full $N$-scan (the count sub-pass), reducing the total from $(I{+}2)$ to
$(I{+}1)$ global-memory passes. Compared to the radix-select baseline which requires $R
\cdot N/P$ accesses with $R \approx 3$--$4$ passes (each pass includes histogram + prefix
sum + filter), the heuristic approach has fewer global memory accesses and significantly
less shared-memory synchronization overhead.

\section{GPU Kernel Implementation on Blackwell}
\label{sec:implementation}

\subsection{Single-CTA Design}
\label{sec:single-cta}

The heuristic Top-K kernel is implemented as a \textbf{single-CTA (Cooperative Thread
Array)} kernel with 512 threads per block. This design enables:
\begin{itemize}
  \item All inter-phase communication via shared memory (no global synchronization).
  \item Efficient reuse of L2 cache across phases (the input data stays warm).
  \item Simple integration as a drop-in replacement for the existing per-row Top-K kernel.
  \item CUDA Graph compatibility (fixed grid dimensions per batch).
\end{itemize}

Each CTA processes one row of the batch (one query token's indexer scores). The multi-row
kernel \texttt{heuristicTopKMultiRowKernel} is a thin wrapper that computes per-row
parameters (sequence length, pointer offsets) and delegates to the
\texttt{heuristicTopKJob} device function.

\subsection{Key Optimizations}
\label{sec:key-optimizations}

The kernel incorporates multiple optimizations targeting specific bottlenecks identified
through systematic profiling:

\begin{table}[t]
  \centering
  \caption{Key kernel optimizations in the GVR Top-K implementation.}
  \label{tab:optimizations}
  \begin{tabular}{@{}ll@{}}
    \toprule
    \textbf{Optimization} & \textbf{Description} \\
    \midrule
    \texttt{\_\_ldg} + \texttt{redux.sync} & Read-only texture cache loads + single-instruction warp reduction \\
    Ballot-free Phase~3 & Eliminates \texttt{\_\_ballot\_sync} barriers that serialize L2 pipelining \\
    Safety-net guard & Skips redundant \texttt{blockCountGE} when Phase~2 converges cleanly \\
    2048-bin histogram & Warp-parallel $K$-th bin search reduces serial scan steps \\
    Phase~3 count-cache & Caches per-thread counts; eliminates redundant $N$-scan \\
    \bottomrule
  \end{tabular}
\end{table}

\subsection{Shared Memory Layout}
\label{sec:smem-layout}

The kernel uses approximately 60\,KB of shared memory:

The layout comprises candidate key/value arrays ($2 \times 6144 \times 4$\,B),
warp counts, a 2048-bin histogram, and the Phase~3 per-thread count
cache---totaling ${\sim}$60\,KB. On Blackwell (sm\_100), the kernel opts in to
extended shared memory ($>$48\,KB) via \texttt{cudaFunc\-Set\-Attribute}. The
\texttt{MAX\_CANDI\-DATES\,=\,6144} buffer provides a 3$\times$ margin over
$K = 2048$, which has been sufficient in the observed decode workloads and reduces the
chance that the safety-net threshold exceeds the refinement capacity. The
full struct definition is provided in Appendix~\ref{app:smem}.

\subsection{Memory Footprint Comparison}
\label{sec:memory-footprint}

The GVR kernel uses ${\sim}$60\,KB SMEM per CTA (${\sim}$2$\times$ the radix-select
baseline's 28\,KB), because candidates must persist from Phase~3 through Phase~4. On
Blackwell (228\,KB per SM), this still allows 3 CTAs per SM. The heuristic path additionally
requires two pre-allocated HBM buffers: a scratch buffer for output values and a per-layer
\texttt{prev\_topk} feedback buffer for temporal prediction. A detailed comparison table and
HBM buffer descriptions are provided in Appendix~\ref{app:smem}.

\subsection{Integration and Activation}
\label{sec:integration}

The heuristic Top-K is integrated into the TensorRT-LLM~\citep{trtllm} DSA pipeline with a
two-level dispatch:

\begin{figure}[t]
  \centering
  \includegraphics[width=0.90\textwidth]{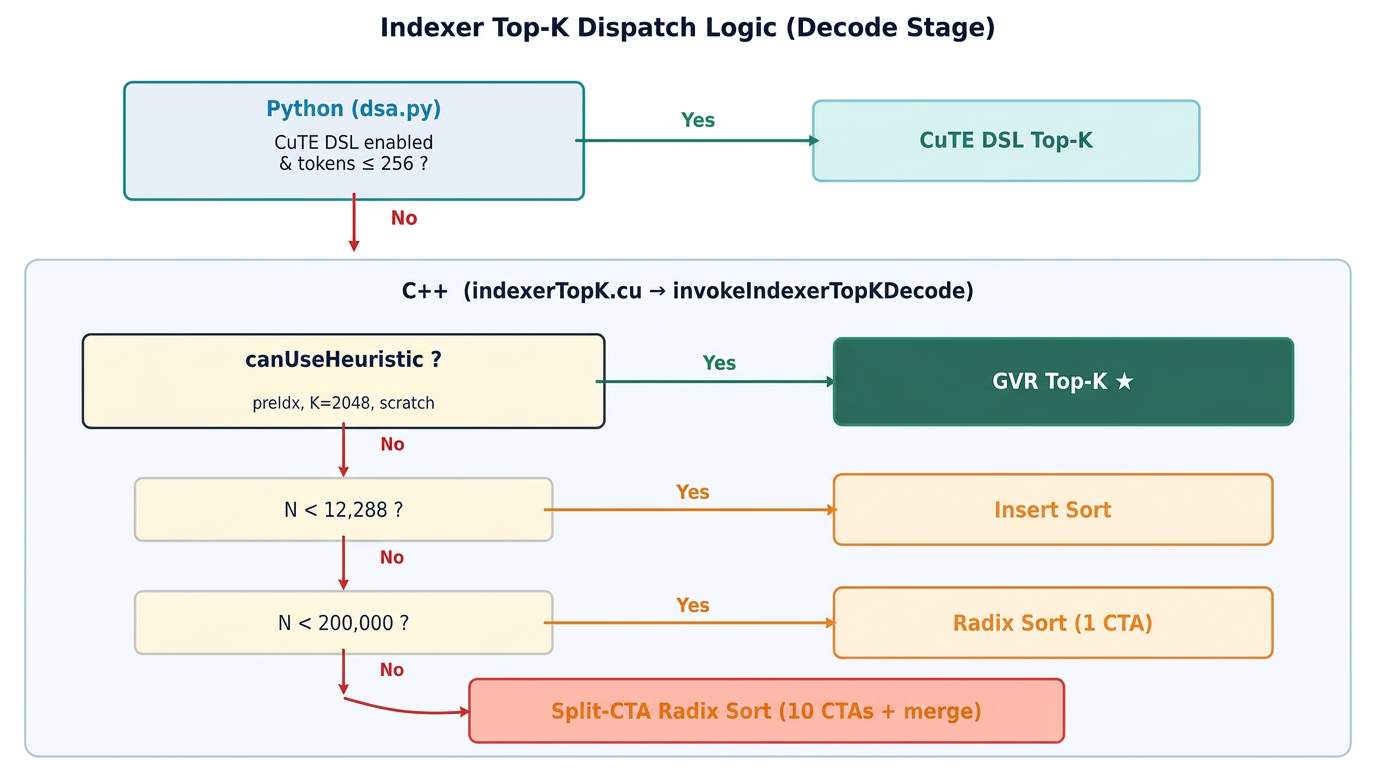}
  \caption{Full decode-stage Top-K dispatch. Python level selects between CuTE DSL Top-K
    (small batch) and the C++ kernel. Within the C++ kernel, the GVR path takes
    highest priority when \texttt{preIdx} and scratch buffer are available; otherwise the
    radix-select fallback chain handles the request.}
  \label{fig:dispatch-logic}
\end{figure}

\textbf{Python level} (\texttt{dsa.py}): If the CuTE DSL Top-K backend is enabled and the
generated-token count is small (\texttt{num\_gen\_tokens} $\leq 256$), it is used directly. Otherwise, the dispatch falls
through to the C++ \texttt{indexer\_topk\_decode} operator, passing the previous step's
Top-K indices (\texttt{pre\_idx}) and a scratch buffer.

\textbf{C++ level} (\texttt{indexerTopK.cu}): A \texttt{canUseHeuristic} gate checks seven
conditions---including non-null \texttt{preIdx}, contiguous layout, $K = 2048$, and
$N < 200$K---before selecting the heuristic fast path. When any condition is not met (e.g.,
prefill phase, first token without \texttt{preIdx}), the dispatch falls through to the
original radix-select pipeline, preserving compatibility with the original path.

\textbf{How to enable.} The heuristic Top-K is controlled by
\texttt{enable\_heuristic\_topk} in the YAML config (default: \texttt{false}), passed via
\texttt{-{}-config} to \texttt{trtllm-serve}, \texttt{trtllm-bench}, or
\texttt{trtllm-eval}. The heuristic path is only active on sm\_100+ (Blackwell) GPUs; on
older architectures the flag is silently ignored. Code listings for the Python dispatch, C++
gate, and YAML configuration are provided in Appendix~\ref{app:integration}.

\section{Evaluation}
\label{sec:evaluation}

\subsection{Experimental Setup}
\label{sec:experimental-setup}

\textbf{Platform}: NVIDIA B200 GPU (Blackwell, sm\_100). All micro-kernel benchmarks are
single-batch, single-row (one CTA), with \textbf{512 threads per block} for both the
heuristic kernel and the production radix-select baseline, ensuring a fair comparison under
identical thread-level resources.

\textbf{Input data} falls into two categories:

\begin{table}[t]
  \centering
  \caption{Input data categories for evaluation.}
  \label{tab:data-categories}
  \begin{tabular}{@{}lllc@{}}
    \toprule
    \textbf{Category} & \textbf{Source} & \textbf{preIdx} & \textbf{Temporal Corr.} \\
    \midrule
    Synthetic random & Random Q/K + YaRN-RoPE & Static RoPE prior & Moderate \\
    Real decode      & DSV3.2 LongSeqTasks logits & Prev-step Top-K  & High \\
    \bottomrule
  \end{tabular}
\end{table}

The synthetic pipeline computes a simplified single-head dot product on RoPE dimensions only
($d_{\mathrm{rope}} = 64$), while the real indexer uses a multi-head weighted sum $I_t =
\sum_j W_j^I \cdot \mathrm{ReLU}(Q_{t,j}^I (K_t^I)^T)$ across 64 heads. Synthetic data
captures positional structure but lacks the content-dependent correlations present in real
decode, resulting in lower \texttt{preIdx} quality and more Phase~2 iterations.

The synthetic data generation pipeline---including YaRN-RoPE frequency computation, static
\texttt{preIdx} construction, and score generation from random Q/K tensors---is detailed
in Appendix~\ref{app:synthetic}.

\textbf{Benchmark methodology}: All kernel timings are collected via \texttt{nsys} profiling
with cold-start input data (L2 cache flushed before each kernel invocation to eliminate
cache-warm artifacts).

\paragraph{Supplementary artifacts.}
To support reproduction of the paper's main results, we release a public
supplementary GitHub repository at
\href{https://github.com/longcheng-nv/GVR_TopK_supplementaty_materials}{this link}.
It contains the public LongSeqTasks prompt buckets, the fixed-OSL TEP8
end-to-end evaluation scripts, the \texttt{nsys}-based timing and
iteration-analysis scripts, and supporting report/figure-generation utilities.
The upstream TensorRT-LLM integration of GVR Top-K is publicly available in the
merged implementation at
\href{https://github.com/NVIDIA/TensorRT-LLM/pull/12385}{TensorRT-LLM PR \#12385}.

\subsection{Correctness Verification}
\label{sec:correctness}

The heuristic Top-K kernel produces \textbf{bit-exact} Top-K index sets compared to
\texttt{torch.topk} across all tested sequence lengths ($N$ = 8K--131K). The output is
non-deterministic for tied values (same as the production kernel); this has negligible
accuracy impact.

\subsection{Single-Operator Performance}
\label{sec:single-op}

\subsubsection{Synthetic Data}
\label{sec:synthetic-data}

\begin{figure}[t]
  \centering
  \includegraphics[width=0.88\textwidth]{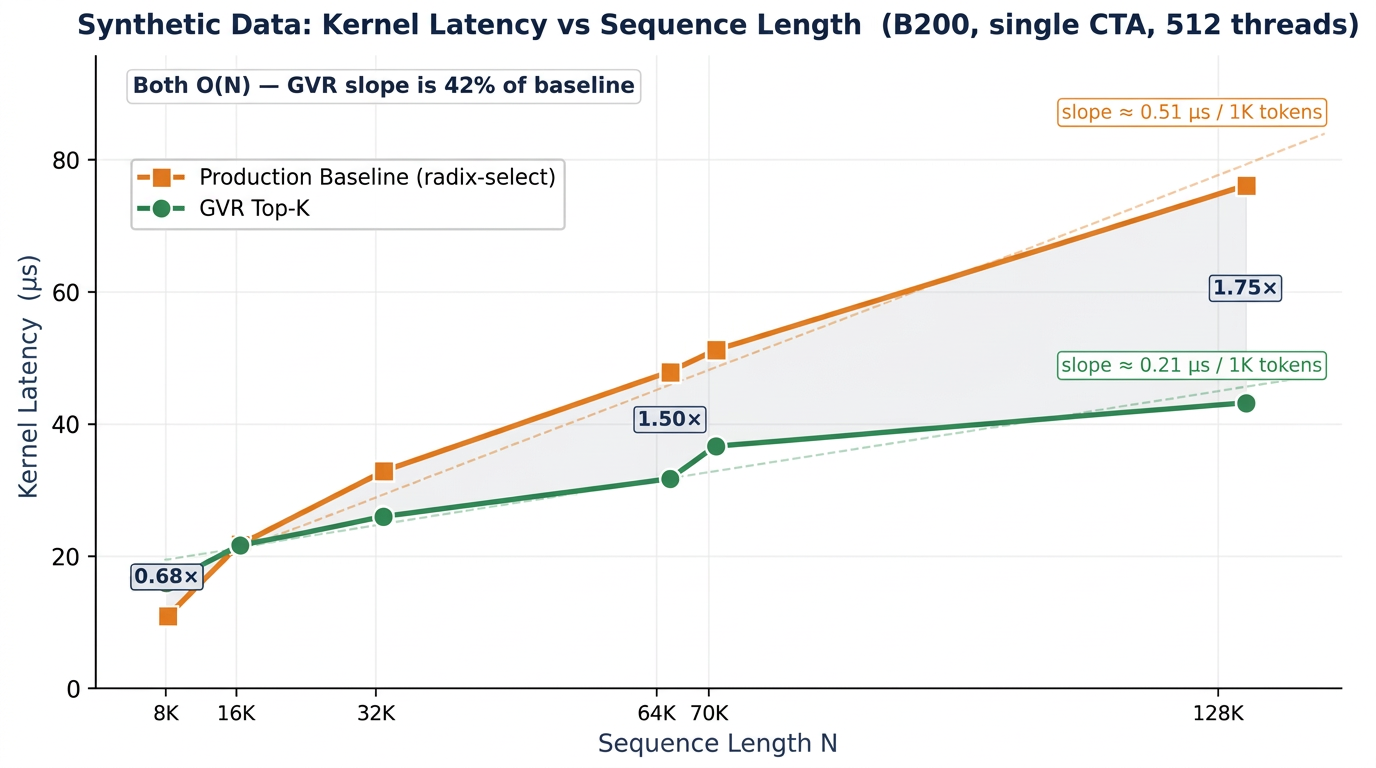}
  \caption{Kernel latency vs.\ sequence length on synthetic data (B200). Both kernels
    scale as $O(N)$; the GVR kernel's fitted slope is ${\sim}$42\% of the baseline,
    reflecting fewer global-memory passes. Speedup labels show the ratio at representative
    $N$ values.}
  \label{fig:synthetic-scaling}
\end{figure}

\begin{table}[t]
  \centering
  \caption{Single-operator latency on B200 with synthetic input. ``Production Baseline'' is
    the \texttt{topKPerRowDecode} kernel (insert sort for $N < 12288$, radix sort for
    $N \geq 12288$). Speedup = baseline time / GVR time.}
  \label{tab:synthetic-latency}
  \begin{tabular}{@{}rccc@{}}
    \toprule
    $N$ & \textbf{Heuristic (ns)} & \textbf{Baseline (ns)} & \textbf{Speedup} \\
    \midrule
      8,192  & 16,512 & 11,200 & 0.68$\times$ \\
     16,384  & 21,856 & 21,984 & \textbf{1.01$\times$} \\
     32,768  & 26,112 & 32,928 & \textbf{1.26$\times$} \\
     65,536  & 31,904 & 47,936 & \textbf{1.50$\times$} \\
     70,690  & 36,864 & 51,200 & \textbf{1.39$\times$} \\
    131,072  & 43,392 & 76,128 & \textbf{1.75$\times$} \\
    \bottomrule
  \end{tabular}
\end{table}

At short sequences ($N = 8192$), the overhead of Phase~1 (scattered \texttt{preIdx} reads)
and Phase~2 (interpolation iterations) outweighs the savings, making the heuristic kernel
${\sim}$32\% slower. The heuristic kernel breaks even around $N = 16384$ and increasingly
outperforms the baseline as sequence length grows---reaching \textbf{1.75$\times$} at $N =
131072$ (Table~\ref{tab:synthetic-latency}). This scaling advantage arises because the
heuristic kernel's global-memory pass count ($I + 1 \approx 3$--$4$) grows slowly relative
to the radix-select approach, whose multi-pass histogram + prefix-sum + filter pipeline
incurs higher per-pass overhead at large~$N$.

\textbf{Key insight}: The static RoPE structural prior used as \texttt{preIdx} in the
synthetic benchmark achieves substantial overlap with the true Top-K, enabling Phase~2 to
converge in fewer iterations than a blind search. Combined with the efficient single-CTA
design and ballot-free collection, this yields a consistent scaling advantage at longer
sequence lengths.

\subsubsection{Real Decoding Data}
\label{sec:real-data}

We evaluate on real DeepSeek-V3.2 decode-stage indexer logits captured from entry~1 of the
64K bucket in SWE-bench-derived LongSeqTasks. This curated long-context
coding prompt is constructed from repository files of a single SWE-bench task; its prompt
length is 68{,}665 tokens and the decode length is 2{,}025 tokens. From the 2{,}024 decode steps
(step~0 skipped as no \texttt{preIdx} is available), 16 evenly spaced samples (stride~128,
starting from step~1) plus the final decode step (step~2{,}024, $N = 70{,}690$) are
selected---17 samples total---for profiling across 9 representative layers. The
\texttt{preIdx} for each sample is the Top-K output from the preceding decode step,
reflecting realistic temporal correlation.

\begin{figure}[t]
  \centering
  \includegraphics[width=0.90\textwidth]{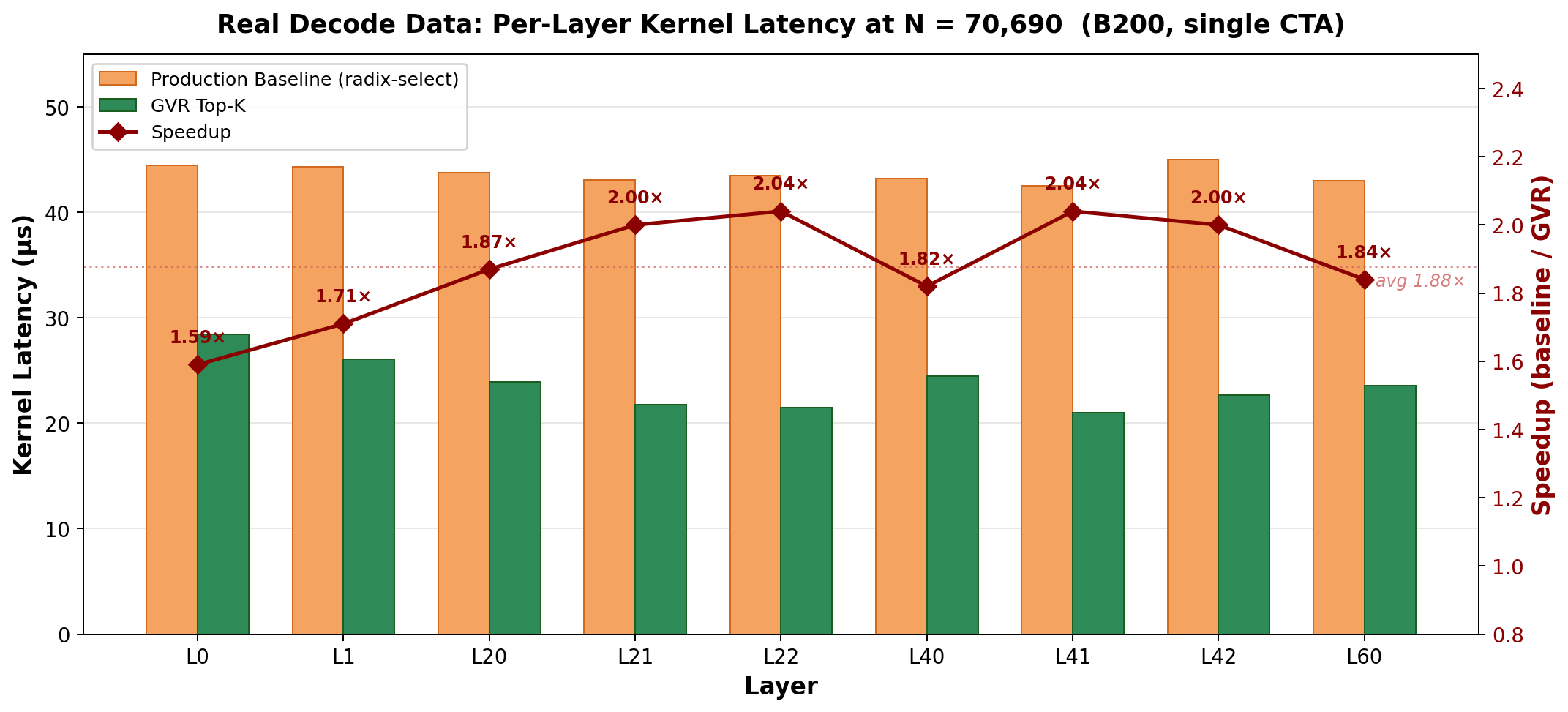}
  \caption{Per-layer kernel latency (averaged over 17 decode steps) at $N \approx 70{,}690$
    on B200. The GVR kernel achieves 1.59$\times$--2.04$\times$ speedup vs.\ the
    production radix-select baseline across all 9 layers (avg 1.88$\times$). L22 and L41
    benefit most (2.04$\times$) due to consistent beta/weibull distributions; L0 benefits
    least (1.59$\times$) due to heterogeneous lognormal distribution.}
  \label{fig:real-data-bars}
\end{figure}

\begin{table}[t]
  \centering
  \caption{Per-layer average, min, and max speedup ratios (radix-select / GVR) across
    17 sampled decode steps (stride 128 from 2{,}024 total). Overall average: 1.88$\times$.}
  \label{tab:real-speedup}
  \begin{tabular}{@{}lccc@{}}
    \toprule
    \textbf{Layer} & \textbf{Avg Speedup} & \textbf{Min Speedup} & \textbf{Max Speedup} \\
    \midrule
     0  & \textbf{1.59$\times$} & 1.15$\times$ & 1.78$\times$ \\
     1  & \textbf{1.71$\times$} & 1.30$\times$ & 1.90$\times$ \\
    20  & \textbf{1.87$\times$} & 1.40$\times$ & 2.21$\times$ \\
    21  & \textbf{2.00$\times$} & 1.48$\times$ & 2.29$\times$ \\
    22  & \textbf{2.04$\times$} & 1.73$\times$ & 2.42$\times$ \\
    40  & \textbf{1.82$\times$} & 1.19$\times$ & 2.23$\times$ \\
    41  & \textbf{2.04$\times$} & 1.74$\times$ & 2.29$\times$ \\
    42  & \textbf{2.00$\times$} & 1.56$\times$ & 2.32$\times$ \\
    60  & \textbf{1.84$\times$} & 1.49$\times$ & 2.18$\times$ \\
    \midrule
    \textbf{Overall} & \textbf{1.88$\times$} & --- & --- \\
    \bottomrule
  \end{tabular}
\end{table}

The results show that the heuristic kernel \textbf{consistently outperforms} the production
radix-select baseline across all layers and all decoding steps. The speedup is remarkably
stable: even the worst-case layer (Layer~0, avg 1.59$\times$) still provides meaningful
improvement, while the best layers (22, 41) achieve 2.04$\times$ average speedup.

On real decoding data, \texttt{pmean} closely approximates the true threshold due to the
high temporal correlation of \texttt{preIdx} (${\sim}$50\% overlap between consecutive
steps), enabling Phase~2 convergence in 1--2 iterations. The variance across layers (e.g.,
Layer~0 having lower speedup than Layer~21) reflects differences in the score distribution
characteristics of individual attention layers---some layers exhibit sharper score
distributions that converge faster, while others have flatter distributions requiring
additional interpolation steps.

\subsection{Data Distribution Analysis and Sensitivity}
\label{sec:distribution-analysis}

Understanding the score distribution per layer is essential for interpreting performance
variance. We performed statistical fitting on the LongSeqTasks decode logits (9 layers
$\times$ 2{,}024 decode steps $\times$ ${\sim}$70{,}690 elements) to characterize each
layer's distribution:

Statistical fitting reveals that \textbf{beta distributions dominate} the deeper layers
(L20--42), enabling fast Phase~2 convergence due to bounded, peaked score distributions.
L0 exhibits a heterogeneous three-way lognorm/beta/weibull split---explaining its
consistently lower speedup---while L1 shows leptokurtic logistic/t behavior with heavier
tails. The per-layer distribution type directly correlates with GVR speedup: bounded
distributions yield ${\geq}$1.82$\times$, while heterogeneous distributions achieve
${\sim}$1.59$\times$. Detailed per-layer fitting results are provided in
Table~\ref{tab:distribution-analysis} in Appendix~\ref{app:distribution-fitting}.

\paragraph{Distribution sensitivity.}
The heuristic Top-K algorithm's speedup is fundamentally sensitive to the input score
distribution, because the distribution determines two key factors: (1)~how close
\texttt{pmean} is to the true $K$-th value (Phase~2 initial guess quality), and (2)~how many
candidates fall near the threshold (Phase~4 snap iteration count).

Combining the synthetic and real-data benchmarks reveals a clear pattern:

\begin{table}[t]
  \centering
  \caption{Distribution sensitivity: GVR speedup as a function of score distribution type.}
  \label{tab:distribution-sensitivity}
  \begin{tabular}{@{}llccc@{}}
    \toprule
    \textbf{Distribution Type} & \textbf{Representative} & \textbf{preIdx Overlap} & \textbf{P2 Iters} & \textbf{Speedup} \\
    \midrule
    Beta (bounded, peaked)     & L21, L40, L41   & High          & 1--2 & \textbf{1.82--2.04$\times$} \\
    Weibull (right-skewed)     & L22, L60        & Moderate--High & 2--3 & \textbf{1.84--2.04$\times$} \\
    Logistic/t (heavy-tailed)  & L1              & Moderate      & 2--3 & \textbf{1.71$\times$} \\
    Lognorm (heterogeneous)    & L0              & Lower         & 3--4 & \textbf{1.59$\times$} \\
    Synthetic (static preIdx)  & $N{=}$70K       & Moderate      & 2--4 & \textbf{1.39$\times$} \\
    \bottomrule
  \end{tabular}
\end{table}

The pattern is consistent for \textbf{Phase~2}: \textbf{bounded, peaked
distributions} (beta) yield the best speedup because \texttt{pmean} sits close
to the Top-K threshold and the candidate count drops quickly during
interpolation. \textbf{Heavy-tailed or heterogeneous distributions} (logistic,
lognorm) spread the score mass across a wider range, causing \texttt{pmean} to
be a less precise initial estimate and therefore requiring more interpolation
iterations.

Phase~4 follows a different rule: snap cost is governed less by \texttt{preIdx}
overlap and more by the local value distribution near the exact Top-K
boundary. Accordingly, L60 has the heaviest snap tail (4.7 average snap
iterations) despite the strongest temporal correlation, whereas L1 is the
cheapest snap layer (2.2 average) despite weak overlap. This is why
high-overlap layers can still spend a large fraction of total time in Phase~4
once Phase~2 becomes cheap.

Notably, even the worst-case real layer (L0, lognorm, 1.59$\times$) and the synthetic
benchmark with static \texttt{preIdx} (1.39$\times$) still consistently outperform the
baseline. This demonstrates the algorithm's \textbf{strong robustness across diverse score
distributions}---delivering stable speedup regardless of whether the underlying distribution
is bounded (beta), heavy-tailed (logistic), skewed (weibull), or heterogeneous (lognorm).

\subsection{Ablation Study: Prediction Signal Quality}
\label{sec:ablation}

To isolate the contribution of the temporal prediction signal, we compare four
\texttt{preIdx} configurations on real DeepSeek-V3.2 LongSeqTasks decode logits ($N
\approx 70{,}690$, 9 layers $\times$ 17 decode steps, \texttt{nsys} GPU trace profiling
with L2 cache flush):

\begin{table}[t]
  \centering
  \caption{Ablation: effect of prediction signal quality on GVR kernel latency.
    High-correlation layers (L20--60) achieve 1.94$\times$ average speedup, while
    low-correlation layers (L0--1) still achieve 1.65$\times$. Even random indices provide
    1.44$\times$ speedup.}
  \label{tab:ablation}
  \begin{tabular}{@{}lccc@{}}
    \toprule
    \textbf{preIdx Source} & \textbf{Overlap ($\alpha$)} & \textbf{Kernel Latency} & \textbf{vs.\ Radix Baseline} \\
    \midrule
    (a) No preIdx (radix fallback) & 0\%       & 43.7\,$\mu$s & 1.00$\times$ (ref.) \\
    (b) Random indices             & ${\sim}$2.9\%  & 30.9\,$\mu$s & \textbf{1.44$\times$} \\
    (c) Prev-step Top-K (L20--60)  & ${\sim}$44\%   & 22.7\,$\mu$s & \textbf{1.94$\times$} \\
    (d) Prev-step Top-K (L0--1)    & ${\sim}$1.5\%  & 27.2\,$\mu$s & \textbf{1.65$\times$} \\
    \bottomrule
  \end{tabular}
\end{table}

Even random \texttt{preIdx} ($\alpha \approx 3$\%) achieves a 1.44$\times$ speedup over
the radix-select baseline. This is because the GVR algorithm's ballot-free Phase~3 design
and count-cache optimization are inherently more efficient than radix-select's
\texttt{atomicAdd}-based histogram serialization---regardless of prediction quality. The
Phase~1 \texttt{pmean} computed from \emph{any} random sample of input values provides a
better initial threshold estimate than a blind radix decomposition, and the ballot-free
candidate collection avoids the \texttt{\_\_ballot\_sync} compiler barriers that serialize
L2 load pipelining in the baseline. In other words, the GVR kernel's architectural
advantages---fewer synchronization points, pre-computed write offsets, and fused
count-cache---deliver substantial gains even when the temporal prediction signal is
effectively absent.

Real temporal prediction ($\alpha \approx 44$\% for L20--60) adds another
${\sim}$0.50$\times$ speedup on top of the architectural baseline, confirming that
prediction quality directly translates to performance via fewer Phase~2 iterations.
Across the 18{,}207-invocation replay above, 67.6\% of real invocations are
one-shot secant solves and 94.8\% finish within 3 iterations. This concentration
comes primarily from the high-correlation layers: L20--60 reach
\textbf{1.94$\times$} average speedup while averaging only 1.1--1.6 Phase~2
iterations (1.3 on average across the group, median 1). Low-correlation layers
(L0--1, $\alpha \approx 1.5$\%) still achieve 1.65$\times$, but their first
\texttt{blockCountGE} count is typically still ${\sim}$31K--34K, far above the
$[K, C] = [2048, 6144]$ target window, so they average 2.1--2.7 Phase~2
iterations before narrowing into range. Even in this regime, the ballot-free
collection and count-cache optimizations keep the per-iteration cost well below
the radix-select baseline's per-pass cost. In the worst case, the GVR kernel's
minimum observed speedup is 0.98$\times$ (essentially matching the radix
baseline), demonstrating graceful degradation. The gap between high- and
low-correlation layers confirms that prediction quality is the \emph{dominant}
factor explaining the per-layer speedup variance observed in
Table~\ref{tab:real-speedup}.

\paragraph{Per-phase time breakdown.}
To quantify each phase's contribution to total kernel latency, we instrumented
the GVR kernel with \texttt{clock64()} timestamps at phase boundaries and
measured per-phase wall-clock durations on real DeepSeek-V3.2 LongSeqTasks
decode logits ($N \approx 70{,}690$). The measurement uses previous-step Top-K
indices as \texttt{preIdx} and samples 17 decode steps per layer at stride~128.
Each configuration was executed 3 times independently with L2 cache flushing
and warm-up; cross-run variation was ${<}$0.3\%.
Appendix~\ref{app:phase-timing} details the instrumentation methodology.

\begin{table}[t]
  \centering
  \caption{Per-phase wall-clock time breakdown of the GVR kernel on B200
    ($N \approx 70{,}690$, $K = 2{,}048$). Each cell
    shows absolute time ($\mu$s) and percentage of total. Boldface marks the
    dominant phase per layer.}
  \label{tab:phase-breakdown}
  \small
  \setlength{\tabcolsep}{3pt}
  \resizebox{\columnwidth}{!}{%
  \begin{tabular}{@{}lccccc@{}}
    \toprule
    \textbf{Layer} & \textbf{Total ($\mu$s)} & \textbf{P1: PreIdx Stats} & \textbf{P2: Threshold Search} & \textbf{P3: Verify/Collect} & \textbf{P4: Refine} \\
    \midrule
    \multicolumn{6}{@{}l}{\emph{Low-correlation layers ($\alpha \approx 1.5$\%):}} \\
    L0  & 34.3 & 3.9\,(12\%) & \textbf{15.7\,(45\%)} & 5.9\,(18\%) & 8.8\,(26\%) \\
    L1  & 28.7 & 4.0\,(14\%) & \textbf{11.4\,(40\%)} & 6.0\,(21\%) & 7.3\,(25\%) \\
    \midrule
    \multicolumn{6}{@{}l}{\emph{High-correlation layers ($\alpha \approx 44$\%):}} \\
    L20 & 27.4 & 2.3\,(9\%)  & \textbf{11.1\,(40\%)} & 5.9\,(22\%) & 8.0\,(30\%) \\
    L21 & 23.4 & 2.2\,(10\%) & 7.5\,(32\%) & 5.9\,(26\%) & \textbf{7.6\,(33\%)} \\
    L22 & 24.4 & 2.7\,(11\%) & 7.0\,(29\%) & 5.9\,(25\%) & \textbf{8.8\,(35\%)} \\
    L40 & 26.5 & 2.7\,(10\%) & \textbf{9.3\,(34\%)} & 5.9\,(23\%) & 8.6\,(33\%) \\
    L41 & 25.3 & 2.5\,(10\%) & 7.7\,(30\%) & 6.0\,(24\%) & \textbf{9.2\,(36\%)} \\
    L42 & 24.4 & 2.3\,(10\%) & \textbf{8.2\,(33\%)} & 6.0\,(25\%) & 7.9\,(32\%) \\
    L60 & 25.4 & 2.5\,(10\%) & \textbf{9.1\,(35\%)} & 5.9\,(24\%) & 7.9\,(32\%) \\
    \bottomrule
  \end{tabular}
  }
\end{table}

The breakdown reveals three structural insights:
\begin{enumerate}
\item \textbf{Phase~3 is bandwidth-bound and layer-invariant.}
  The ballot-free collection phase consumes a near-constant $5.9$--$6.0\,\mu$s
  across all 9 layers, confirming that it is limited by single-pass HBM read
  bandwidth rather than by computation or prediction quality. This validates
  the streaming single-CTA design: Phase~3 performs exactly one sequential scan
  regardless of the threshold search outcome.

\item \textbf{Phase~2 drives inter-layer variance.}
  Threshold search latency ranges from $7.0\,\mu$s (L22, 29\%) to
  $15.7\,\mu$s (L0, 45\%)---a $2.2\times$ span that closely mirrors the
  per-layer speedup spread in Table~\ref{tab:real-speedup}. The replay-based
  control-flow statistics show that 94.8\% of real invocations finish Phase~2
  within 3 iterations, but the layer-wise averages still separate cleanly by
  correlation regime: low-correlation layers (L0--1) average 2.1--2.7 secant
  iterations (median 2, max 5) because \texttt{pmean} from
  $\alpha \approx 1.5$\% overlap is a poor initial estimate, whereas
  high-correlation layers (L21--60) average 1.1--1.6 iterations (median 1).
  L20, despite high overlap, shows elevated Phase~2 cost ($11.1\,\mu$s) due
  to its heterogeneous beta/weibull split
  (Table~\ref{tab:distribution-analysis}).

\item \textbf{Phase~4 becomes the dominant cost for high-correlation layers.}
  When Phase~2 converges quickly, histogram exact selection emerges as the
  bottleneck at $7.3$--$9.2\,\mu$s (32--36\% of total). This is
  ${\sim}1.7\times$ higher than the initial analytical estimate of
  $4$--$5\,\mu$s, attributable to the histogram-sort refinement step that
  resolves ties among candidates sharing the threshold value. The replay-based
  snap statistics show that this cost is not predicted by overlap alone:
  L60 has the heaviest snap tail (4.7 average, P99 12.8) despite the strongest
  temporal correlation, whereas L1 averages only 2.2 snap iterations. In short,
  Phase~4 cost is driven primarily by the local score geometry near the exact
  Top-K boundary rather than by prediction quality itself.
\end{enumerate}

\paragraph{Optimization opportunities.}
The phase breakdown identifies two distinct optimization targets by layer
regime. For high-correlation layers (L21--60), Phase~4 accounts for up to
36\% of total kernel time; potential improvements include fused
histogram-scatter operations to reduce Phase~4 memory traffic, or adaptive
bin-width strategies that exploit the known distribution shape
(Section~\ref{sec:distribution-analysis}) to reduce snap/tie-breaking
iterations near the exact threshold.
For low-correlation layers (L0--1), reducing Phase~2 iteration count via
distribution-aware initial threshold estimation would yield the largest gain.
Phase~1 accounts for only 10--14\% of total time, confirming that the
\texttt{pmean} pre-computation is efficiently amortized.

\subsection{End-to-End Accuracy}
\label{sec:e2e-accuracy}

We next validate that enabling GVR does not introduce measurable end-to-end quality
regression in the full TensorRT-LLM stack. We use \texttt{trtllm-eval} on four
benchmarks---MMLU~\citep{mmlu}, GSM8K~\citep{gsm8k},
GPQA-Diamond~\citep{gpqa}, and LongBench~\citep{longbench}---with
DeepSeek-V3.2 NVFP4 on B200, and run each benchmark multiple times
independently to assess run-to-run variance. These results should be interpreted as
end-to-end regression coverage for the deployment stack rather than as equally strong stress
tests of the decode-stage Top-K path: the long-context LongBench setting is the most
representative of sustained decode-time DSA usage, while short-output benchmarks primarily
verify that enabling the heuristic path does not perturb model behavior outside the
long-context regime.

\begin{table}[t]
  \centering
  \caption{End-to-end accuracy on DeepSeek-V3.2 NVFP4. ``Exps (B/H)'' = independent
    runs for Baseline / Heuristic.}
  \label{tab:e2e-accuracy}
  \scriptsize
  \begin{tabular}{@{}lrcccc@{}}
    \toprule
    \textbf{Benchmark} & \textbf{Samples} & \textbf{Baseline (avg)} & \textbf{Heuristic (avg)} & \textbf{Delta} & \textbf{Exps (B/H)} \\
    \midrule
    MMLU           & 14,042  & 87.51 & 87.50 & $\mathbf{-0.01}$ & 1 / 4 \\
    GSM8K          &  1,319  & 95.11 & 95.23 & $\mathbf{+0.12}$ & 1 / 4 \\
    GPQA-Diamond   &    198  & 77.27 & 77.15 & $\mathbf{-0.12}$ & 1 / 4 \\
    LongBench V1   & ${\sim}$5,000 & 44.61 & 44.28 & $\mathbf{-0.33}$ & 8 / 8 \\
    \bottomrule
  \end{tabular}
\end{table}

\subsection{End-to-End Min-Latency Scaling with Context Length}
\label{sec:e2e-latency}

\begin{figure}[!t]
  \centering
  \includegraphics[width=0.80\columnwidth]{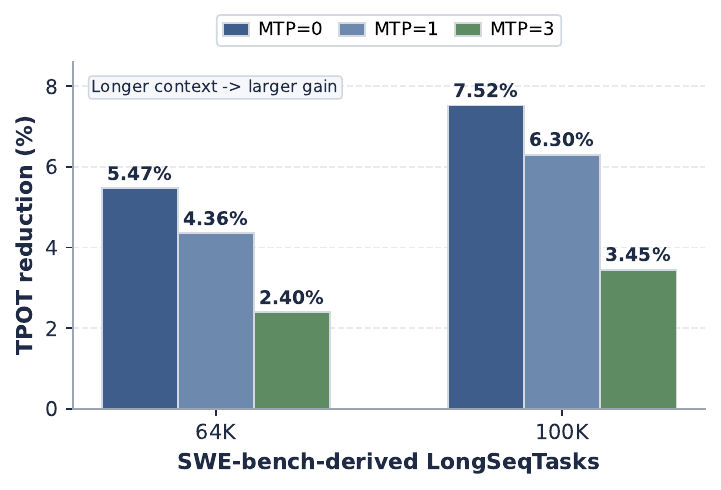}
  \caption{End-to-end TPOT reduction at fixed OSL=1K on DeepSeek-V3.2-Exp NVFP4 with
    TensorRT-LLM on B200$\times$8 under TEP8 parallelism. Colors denote the speculative
    setting (MTP=0/1/3). GVR gains increase with context length and are largest at MTP=0.}
  \label{fig:e2e_tept8_osl1k_bar}
\end{figure}

We next evaluate fixed-OSL=1K min-latency inference under controlled TEP8 settings on
DeepSeek-V3.2-Exp NVFP4 with TensorRT-LLM on B200$\times$8, using a fixed LongSeqTasks
prompt and 3 alternating A/B repetitions of the production radix-select and GVR Top-K paths.
Figure~\ref{fig:e2e_tept8_osl1k_bar} shows a clear length-scaling trend: without speculative
decoding (MTP=0), the TPOT reduction rises from \textbf{5.47\%} at 64K to
\textbf{7.52\%} at 100K, and the same trend holds under MTP=1 and MTP=3.

The three bars at each context length correspond to MTP=0, MTP=1, and MTP=3. Here ``64K''
and ``100K'' refer to the two public LongSeqTasks buckets,
\texttt{swe\_bench\_64k.jsonl} and \texttt{swe\_bench\_100k.jsonl}; in both cases we fix
entry~1 to control workload variation across A/B repetitions. Under MTP=1, the reduction
rises from \textbf{4.36\%} to \textbf{6.30\%}, while under MTP=3 the gain is smaller but
follows the same trend, improving from \textbf{2.40\%} to \textbf{3.45\%}. This is
consistent with the core mechanism of GVR: as sequence length increases, the Top-K selector
occupies a larger fraction of the decode critical path, making the reduction in Top-K
overhead increasingly valuable.

The monotonic ordering MTP=0 $>$ MTP=1 $>$ MTP=3 is also expected from Amdahl's law:
speculative decoding amortizes the fixed per-step Top-K savings across multiple output
tokens, so the end-to-end benefit of optimizing Top-K is diluted as MTP increases. Even so,
the context-length trend remains the same across all three settings. Importantly, the draft
acceptance rate remains unchanged within noise for the speculative cases, indicating that the
heuristic Top-K path introduces no measurable end-to-end accuracy degradation. Detailed
per-repetition TPOT values at OSL=1K are provided in Table~\ref{tab:e2e_tept8_osl1k_detail},
and the corresponding DAR values for MTP$>0$ are listed in
Table~\ref{tab:e2e_tept8_osl1k_dar} in the appendix.
We intentionally keep broader synthetic stress tests and throughput/Pareto exploration out of
the main text, since the fixed-OSL real-prompt benchmark above more directly reflects the
context-length scaling trend relevant to min-latency serving, while broader serving-region
behavior is more sensitive to
batching and scheduling policy than the controlled batch-1 min-latency setting considered
here.

\section{Related Work}
\label{sec:related-work}

\paragraph{GPU Top-K selection.}
Efficient Top-K selection on GPUs is well studied. \citet{zhangsc23} benchmark radix,
sampling, bucket-sort, and bitonic-sort variants across a wide $(N, K)$ space on A100 and
H100, showing that the best algorithm depends on regime, data distribution, and GPU
microarchitecture; this directly motivates workload-adaptive designs. RadiK~\citep{radik}
further optimizes radix-select for Ampere and Hopper, though no Blackwell kernel is publicly
available. \citet{zois2019topk} propose early stopping to reduce average-case passes, while
\citet{approxtopk} study approximate Top-K that trades exactness for parallelism. These
methods treat Top-K as a \emph{distribution-agnostic} primitive; our work instead exploits
\emph{temporal correlation} in autoregressive decoding to provide a data-dependent warm-start
while retaining exact correctness.

\paragraph{Sparse attention for long-context LLMs.}
As context lengths grow into the 100K+ range, sparse attention has become a practical
strategy for tractable inference. DSA~\citep{deepseekv32, deepseekv3} uses a lightweight
indexer with Top-K selection to identify the $K$ most important tokens per query. Orthogonal
DSA optimizations include hierarchical indexers that prune the search space before
token-level refinement~\citep{dsaorth2026}. NSA~\citep{nsa}, MoBA~\citep{moba},
RocketKV~\citep{rocketkv}, Quest~\citep{quest}, and SAGE-KV~\citep{sagekv} explore
block-sparse routing, KV compression, query-aware sparsity, and cache-management variants.
More recent DeepSeek long-context architectures such as DeepSeek-V4 combine CSA and HCA
while retaining DSA within the CSA branch, suggesting that indexer-guided sparse selection
remains relevant even as the downstream sparse-attention mechanism evolves~\citep{deepseekv4}.
Across these methods, Top-K selection at token, page, or block granularity remains a
critical primitive. GVR should apply whenever decode-stage Top-K exhibits temporal
correlation, although we validate it only on DSA in this work.

\paragraph{Positional encodings and temporal structure.}
RoPE~\citep{rope} encodes position through rotation matrices whose frequency structure
produces the Toeplitz property exploited in this work. YaRN~\citep{yarn} extends RoPE to
longer contexts by interpolating low-frequency components, which---as we show---preserves
significant peaks at large relative positions and enhances the temporal correlation signal
available for Top-K prediction. The connection between RoPE frequency structure and Top-K
temporal stability appears to be novel to this work.

\section{Conclusion and Future Work}
\label{sec:conclusion}

We presented Guess-Verify-Refine (GVR), a data-aware exact Top-K algorithm for
sparse-attention decoding on NVIDIA Blackwell GPUs. By using the previous step's Top-K as a
prediction signal, GVR combines secant-style threshold search, ballot-free candidate
collection, and in-shared-memory exact refinement to reduce full-row passes from 3--4 to 1--2
without sacrificing exactness.

On real DeepSeek-V3.2 decode workloads, GVR achieves an average
\textbf{1.88$\times$} single-operator speedup, with up to \textbf{2.42$\times$} per layer
per step. In controlled TEP8 min-latency deployment, it preserves model quality and reduces
end-to-end TPOT by up to \textbf{7.52\%}, with larger gains at longer contexts and smaller
but still positive gains under speculative decoding. Together with the replay-based Phase~2
statistics and prediction-quality ablation, these results show that workload-aware prediction
can substantially improve exact Top-K in the evaluated TensorRT-LLM sparse-attention serving
setting.

\paragraph{Limitations.}
This study focuses on decode-stage exact Top-K for DeepSeek-style sparse attention on
Blackwell, where temporal stability provides a strong prediction signal. The current
implementation is a single-CTA design integrated into the outer TensorRT-LLM indexer Top-K
framework, so it is most effective in long-context decode and can lose benefit at shorter
sequence lengths. Small-batch regimes still leave room for intra-row multi-CTA parallelism to
improve occupancy, while larger-batch regimes would benefit from reducing the current SMEM
footprint through tighter \texttt{MAX\_CANDIDATES} sizing and broader shared-memory
minimization. Exploring those system-level optimizations, together with cross-architecture and
cross-GPU validation, remains future work.

\paragraph{Future directions.}
Promising next steps include adaptive switching for short sequences, extending GVR beyond the
current single-CTA regime to ultra-long and higher-throughput decode settings, developing
stronger signals for prefill and MTP-aware decode, and validating the same principle across
other sparse-attention architectures and GPU generations.

\section*{Acknowledgments}
The authors gratefully acknowledge Christina Zhang, Mengdi Wang, Yuhang He, and Fanrong Li
from NVIDIA for their invaluable support, constructive feedback, and insightful discussions
throughout this work.
The authors also thank other collaborators and supporting team members whose assistance with the
experimental environment, evaluation, and internal review process made this work possible.

\clearpage
\begingroup
\small
\setlength{\bibsep}{1pt plus 0.2ex}
\bibliographystyle{plainnat}
\bibliography{references}
\endgroup

\clearpage
\appendix

\section{End-to-End TPOT at Fixed OSL=1K}
\label{app:e2e-osl1k}

\noindent This appendix section reports the per-repetition TPOT values for MTP=0/1/3 and the
corresponding DAR values for MTP$>0$ underlying the main-text
Figure~\ref{fig:e2e_tept8_osl1k_bar}. The measurements use the same controlled setup as the
main-text latency study: DeepSeek-V3.2-Exp NVFP4 on B200$\times$8 under TEP8, batch size~1,
a fixed LongSeqTasks prompt entry, 5 requests per configuration, and 3 alternating A/B
repetitions.

\begin{table}[H]
  \centering
  \caption{Per-repetition end-to-end TPOT at OSL=1K under the same fixed-OSL TEP8
    setup as Figure~\ref{fig:e2e_tept8_osl1k_bar}. Reduction is computed from the
    mean TPOT values across the three alternating A/B repetitions.}
  \label{tab:e2e_tept8_osl1k_detail}
  \scriptsize
  \setlength{\tabcolsep}{3pt}
  \renewcommand{\arraystretch}{1.05}
  \resizebox{\columnwidth}{!}{%
  \begin{tabular}{@{}llccccccccc@{}}
    \toprule
    \textbf{MTP} & \textbf{ISL}
      & \multicolumn{3}{c}{\textbf{Baseline TPOT (ms)}}
      & \multicolumn{3}{c}{\textbf{GVR TPOT (ms)}}
      & \textbf{Mean Base}
      & \textbf{Mean GVR}
      & \textbf{Reduction} \\
    \cmidrule(lr){3-5} \cmidrule(lr){6-8}
      & & \textbf{R1} & \textbf{R2} & \textbf{R3}
      & \textbf{R1} & \textbf{R2} & \textbf{R3}
      & & & \\
    \midrule
    0 & 64K  & 10.8161 & 10.8509 & 10.8785 & 10.2708 & 10.2367 & 10.2593 & 10.8485 & 10.2556 & \textbf{5.47\%} \\
    0 & 100K & 11.4964 & 11.5023 & 11.5316 & 10.6522 & 10.6438 & 10.6359 & 11.5101 & 10.6440 & \textbf{7.52\%} \\
    \midrule
    1 & 64K  & 6.5508 & 6.5188 & 6.5032 & 6.2436 & 6.2207 & 6.2561 & 6.5243 & 6.2401 & \textbf{4.36\%} \\
    1 & 100K & 7.4759 & 7.6306 & 7.6171 & 7.0760 & 7.1353 & 7.0796 & 7.5745 & 7.0970 & \textbf{6.30\%} \\
    \midrule
    3 & 64K  & 5.1900 & 5.2674 & 5.1760 & 5.0277 & 5.1663 & 5.0635 & 5.2111 & 5.0858 & \textbf{2.40\%} \\
    3 & 100K & 5.5394 & 5.5781 & 5.7050 & 5.5157 & 5.3746 & 5.3520 & 5.6075 & 5.4141 & \textbf{3.45\%} \\
    \bottomrule
  \end{tabular}
  }
\end{table}

\begin{table}[H]
  \centering
  \caption{Per-repetition draft acceptance rate (DAR) at OSL=1K for the speculative
    settings (MTP$>0$) under the same fixed-OSL TEP8 setup. The small differences remain
    within the run-to-run variation of speculative decoding.}
  \label{tab:e2e_tept8_osl1k_dar}
  \small
  \setlength{\tabcolsep}{3pt}
  \renewcommand{\arraystretch}{1.05}
  \makebox[\columnwidth][c]{%
  \begin{tabular}{@{}llccccccccc@{}}
    \toprule
    \textbf{MTP} & \textbf{ISL}
      & \multicolumn{3}{c}{\textbf{Baseline DAR}}
      & \multicolumn{3}{c}{\textbf{GVR DAR}}
      & \textbf{Mean Base}
      & \textbf{Mean GVR}
      & \textbf{$\Delta$DAR} \\
    \cmidrule(lr){3-5} \cmidrule(lr){6-8}
      & & \textbf{R1} & \textbf{R2} & \textbf{R3}
      & \textbf{R1} & \textbf{R2} & \textbf{R3}
      & & & \\
    \midrule
    1 & 64K  & 0.91 & 0.93 & 0.93 & 0.92 & 0.93 & 0.93 & 0.92 & 0.93 & +0.01 \\
    1 & 100K & 0.91 & 0.87 & 0.88 & 0.89 & 0.87 & 0.90 & 0.89 & 0.89 & +0.00 \\
    \midrule
    3 & 64K  & 0.61 & 0.60 & 0.62 & 0.61 & 0.61 & 0.62 & 0.61 & 0.61 & +0.00 \\
    3 & 100K & 0.58 & 0.60 & 0.58 & 0.57 & 0.59 & 0.58 & 0.59 & 0.58 & -0.01 \\
    \bottomrule
  \end{tabular}
  }
\end{table}

\section{Pre-Computed Candidate Index Analysis}
\label{app:preidx-analysis}

An earlier approach considered using \textbf{peak indices} of $g(\Delta)$ (positions where
$g'(\Delta) = 0$ and $g''(\Delta) < 0$) as candidates. However, analysis showed that peak
indices achieve only ${\sim}$17--35\% prediction success rate---the ratio of significant
peaks to total tokens is low (${\sim}$13\%), and many true Top-K indices fall between peaks.
The \textbf{TopK-based prediction} (using the Top-K of $g(\Delta)$ directly) achieves
45--100\% success rate, a several-fold improvement.

\section{Shared Memory Layout and Memory Footprint Details}
\label{app:smem}

\noindent\begin{minipage}{\linewidth}
\begin{lstlisting}[language=C++, caption={Shared memory layout of the GVR Top-K kernel.},
  label={lst:smem}]
KernelSmem {
    float  keys[MAX_CANDIDATES];          // 6144 x 4B = 24,576 B
    int    vals[MAX_CANDIDATES];          // 6144 x 4B = 24,576 B
    int    warp_counts[NUM_WARPS];        //   16 x 4B =     64 B
    int    histogram[NUM_BINS];           // 2048 x 4B =  8,192 B
    int    per_thread_counts[BLOCK_SIZE]; //  512 x 4B =  2,048 B  // P3 count cache
    // + scalar temporaries                             ~    40 B
};                                        // Total: ~60 KB
\end{lstlisting}
\end{minipage}\par

\begin{table}[H]
  \centering
  \caption{Memory footprint comparison between radix-select and GVR Top-K.}
  \label{tab:memory-footprint}
  \begin{tabular}{@{}lcc@{}}
    \toprule
    \textbf{Metric} & \textbf{Radix Select} & \textbf{GVR (Heuristic)} \\
    \midrule
    SMEM per CTA       & ${\sim}$28\,KB (union, time-multiplexed) & ${\sim}$60\,KB (flat, persistent) \\
    Extended SMEM opt-in & No                                     & Yes ($>$48\,KB) \\
    Additional HBM       & 0                                      & Scratch + prev\_topk \\
    \bottomrule
  \end{tabular}
\end{table}

The heuristic path requires two additional persistently pre-allocated HBM buffers (CUDA
Graph compatible):
\begin{itemize}
  \item \texttt{heuristic\_scratch\_values} ($B_{\mathrm{batch}} \times K \times 4$ bytes,
    where $B_{\mathrm{batch}}$ is the batch size): A write-only
    dummy buffer for the kernel's output-values path. The DSA pipeline only consumes the
    output indices, but the kernel unconditionally writes values alongside indices to
    preserve optimal SASS code generation quality.
  \item \texttt{heuristic\_prev\_topk} ($L \times B_{\mathrm{batch}} \times K \times 4$
    bytes, where
    $L$ = number of local DSA layers): Stores the previous decode step's Top-K indices per
    layer as \texttt{preIdx} for the next step. A dedicated buffer is required because
    (1)~\texttt{topk\_indices\_buffer} is overwritten in-place each step,
    (2)~each layer needs independent temporal state, and
    (3)~CUDA Graph replay requires stable-address feedback buffers.
\end{itemize}

\section{Integration Code Listings}
\label{app:integration}

\textbf{Python-level dispatch} (\texttt{dsa.py}):

\noindent\begin{minipage}{\linewidth}
\begin{lstlisting}[language=Python, caption={Python-level dispatch in \texttt{dsa.py}.},
  label={lst:python-dispatch}]
if self.use_cute_dsl_topk and num_gen_tokens <= 256:
    torch.ops.trtllm.cute_dsl_indexer_topk_decode(
        logits, kv_lens, indices, topk, next_n)
else:
    torch.ops.trtllm.indexer_topk_decode(
        logits, kv_lens, indices, next_n, topk,
        pre_idx=pre_idx,               # previous step's Top-K
        heuristic_scratch=heuristic_scratch)  # scratch buffer
\end{lstlisting}
\end{minipage}\par

\textbf{C++ heuristic gate} (\texttt{indexerTopK.cu}):

\noindent\begin{minipage}{\linewidth}
\begin{lstlisting}[language=C++, caption={C++ GVR dispatch gate in \texttt{indexerTopK.cu}.},
  label={lst:cpp-gate}]
bool const canUseHeuristic = preIdx != nullptr
    && stride1 == 1                       // contiguous memory layout
    && topK == kHeuristicTopK             // K = 2048
    && preIdxCount == kHeuristicSize      // M = 2048
    && preIdxStride >= preIdxCount        // valid stride
    && numColumns < effectiveSplitWorkThreshold  // N < 200K (single-CTA)
    && heuristicScratch != nullptr;       // scratch buffer allocated
\end{lstlisting}
\end{minipage}\par

\textbf{YAML configuration}:

\noindent\begin{minipage}{\linewidth}
\begin{lstlisting}[language=yaml, caption={Enabling GVR Top-K in the serving
    configuration.}, label={lst:yaml-config}]
# config.yml (or extra_llm_api_options.yaml)
sparse_attention_config:
    algorithm: dsa
    enable_heuristic_topk: true
\end{lstlisting}
\end{minipage}\par

\section{Synthetic Data Generation}
\label{app:synthetic}

\noindent\begin{minipage}{\linewidth}
\begin{lstlisting}[language=Python, caption={Synthetic indexer score generation with
    YaRN-RoPE frequency computation.}, label={lst:synthetic}]
import math, torch, numpy as np

def yarn_inv_freq(dim=64, base=10000.0, sf=40.0, orig_max=4096, bf=32, bs=1):
    """DeepSeek-V3.2 YaRN frequency computation."""
    pos_f = base ** (torch.arange(0, dim, 2, dtype=torch.float32) / dim)
    freq_extra, freq_inter = 1.0 / pos_f, 1.0 / (sf * pos_f)
    lo = max(int(dim * math.log(orig_max/(bf*2*math.pi))
             / (2*math.log(base))), 0)
    hi = min(int(math.ceil(dim*math.log(orig_max/(bs*2*math.pi))
             / (2*math.log(base)))), dim-1)
    ramp = torch.clamp(
        (torch.arange(dim//2).float() - lo) / max(hi-lo, 1e-3), 0, 1)
    return freq_inter * ramp + freq_extra * (1 - ramp)

def compute_static_pre_idx(N, K=2048, d_rope=64):
    """Compute preIdx from the all-ones RoPE structural prior."""
    theta = yarn_inv_freq(d_rope).numpy()
    f = 2 * np.cos(np.outer(np.arange(N), theta)).sum(axis=1)
    return torch.from_numpy(f).topk(K).indices.int()

def generate_indexer_scores(N, K=2048, Am=0.1, d_rope=64, device="cuda"):
    """Generate synthetic scores (random Q/K + YaRN-RoPE) and static preIdx."""
    inv_freq = yarn_inv_freq(d_rope).to(device)
    pos = torch.arange(N, device=device).float()
    cos_t = torch.cos(pos[:, None] * inv_freq[None, :])
    sin_t = torch.sin(pos[:, None] * inv_freq[None, :])
    def rope(x, c, s):
        x1, x2 = x[..., ::2], x[..., 1::2]
        return torch.cat([x1*c - x2*s, x2*c + x1*s], dim=-1)
    q = 1.0 + Am * torch.randn(1, d_rope, device=device)
    k = 1.0 + Am * torch.randn(N, d_rope, device=device)
    scores = (rope(q, cos_t[:1], sin_t[:1])
              @ rope(k, cos_t, sin_t).T).squeeze(0)
    pre_idx = compute_static_pre_idx(N, K, d_rope).to(device)
    return scores, pre_idx
\end{lstlisting}
\end{minipage}\par

\section{Per-Layer Distribution Fitting Details}
\label{app:distribution-fitting}

\begin{table}[H]
  \centering
  \caption{Per-layer distribution characterization from statistical fitting on LongSeqTasks
    decode logits.}
  \label{tab:distribution-analysis}
  \small
  \begin{tabular}{@{}lcccccc@{}}
    \toprule
    \textbf{Layer} & \textbf{Best-Fit} & \textbf{\% Rows} & \textbf{Runner-Up} & \textbf{\% Rows} & \textbf{Mean Logit} & \textbf{Kurtosis} \\
    \midrule
    L0  & lognorm      & 42.9\% & beta         & 26.0\% & $-4.12$ & $-0.128$ \\
    L1  & logistic     & 59.4\% & t            & 40.6\% & $-0.47$ & $+0.931$ \\
    L20 & beta         & 90.1\% & weibull\_min & 9.9\%  & $-0.65$ & $-0.282$ \\
    L21 & beta         & 99.7\% & weibull\_min & 0.2\%  & $-0.87$ & $-0.337$ \\
    L22 & weibull\_min & 63.9\% & beta         & 36.1\% & $-3.04$ & $-0.111$ \\
    L40 & beta         & 86.1\% & lognorm      & 5.0\%  & $-3.16$ & $-0.162$ \\
    L41 & beta         & 79.5\% & lognorm      & 10.3\% & $-2.76$ & $-0.138$ \\
    L42 & beta         & 63.7\% & weibull\_min & 36.4\% & $-4.51$ & $-0.621$ \\
    L60 & weibull\_min & 93.6\% & beta         & 6.3\%  & $-2.26$ & $-0.396$ \\
    \bottomrule
  \end{tabular}
\end{table}

Key observations for threshold search performance:
\begin{itemize}
  \item \textbf{Beta distributions dominate} L20/21/40/41/42---bounded, shaped distributions
    where \texttt{pmean} accurately approximates the Top-K threshold, enabling fast Phase~2
    convergence (1--2 iterations).
  \item \textbf{L1 is leptokurtic} (kurtosis $+0.931$) with heavy tails (logistic/t
    split)---harder threshold estimation, more Phase~2 iterations.
  \item \textbf{L0 is the most heterogeneous} (three-way split: lognorm/beta/weibull) with
    the widest value range (17.28)---explains its consistently lower speedup across
    benchmarks.
  \item All mean logit values are negative ($-0.47$ to $-4.51$), consistent with
    post-attention indexer logit space.
\end{itemize}

\section{Per-Benchmark Accuracy Details}
\label{app:accuracy-details}

\begin{itemize}
  \item \textbf{MMLU} (14{,}042 questions, single-token output): GVR runs
    are deterministic at 87.52\slash 87.51\slash 87.49\slash
    87.49---virtually identical to the baseline~87.51. The DSA indexer
    Top-K is not exercised on short-output benchmarks since decode
    sequence lengths are minimal.
  \item \textbf{GSM8K} (1{,}319 math problems): GVR average 95.23 (range
    95.03--95.49) vs.\ baseline 95.11---consistent with the NVFP4 checkpoint
    reference of 95.26.
  \item \textbf{GPQA-Diamond} (198 questions, thinking mode): GVR average 77.15
    (range 75.25--78.79) vs.\ baseline 77.27. The high per-experiment variance
    ($\pm$3 points) is expected given the small sample size; both within noise.
  \item \textbf{LongBench V1} (${\sim}$5{,}000 tasks, avg ISL ${\sim}$10K):
    8 independent experiments each. Baseline range 44.25--45.01, GVR range
    43.73--44.67. The delta ($-0.33$) is within the $\pm$0.76 run-to-run spread.
\end{itemize}

\section{Per-Phase Timing Measurement Methodology}
\label{app:phase-timing}

The per-phase time breakdown in Table~\ref{tab:phase-breakdown} was obtained
using compile-time instrumentation of the GVR kernel. This appendix documents
the measurement methodology and reproducibility assessment.
The host-side scripts for the timing extraction and replay-based iteration
analysis are included in the supplementary artifact repository referenced in
Section~\ref{sec:experimental-setup}.

\paragraph{Instrumentation.}
The GVR kernel supports fine-grained per-phase timing when compiled with the
\texttt{-DGVR\_PHASE\_TIMING} preprocessor flag. Under this flag, the kernel
writes five \texttt{clock64()} timestamps into the scratch buffer at phase
boundaries: $t_0$ (kernel entry), $t_1$ (end of Phase~1), $t_2$ (end of
Phase~2), $t_3$ (end of Phase~3), and $t_4$ (end of Phase~4 / kernel exit).
The timestamps are stored as five 64-bit integers packed into 10 consecutive
\texttt{float32} slots at \texttt{scratch[topK\,:\,topK{+}10]}, piggybacking
on the existing scratch buffer without additional memory allocation. A
host-side script extracts these timestamps after each kernel invocation,
computes cycle deltas $\Delta_i = t_{i+1} - t_i$ for each phase, and converts
to wall-clock microseconds via the SM clock frequency.

\paragraph{Experimental protocol.}
All measurements were conducted on a single NVIDIA B200 GPU with locked SM
clock. Input data consists
of real DeepSeek-V3.2 LongSeqTasks decode logits (9 representative layers
$\times$ 2{,}024 decode steps, $N \approx 68{,}667$--$70{,}690$ per step,
growing by 128 per step as the KV cache extends). For each layer, 17 decode
steps are sampled at stride~128 (steps 1, 129, 257, \ldots, 1921, 2024),
covering the full decode trajectory. Each sampled step uses the previous step's
Top-K indices as \texttt{preIdx}, matching the production runtime
configuration. Before each timed invocation, the L2 cache is flushed via a
128\,MB zero-fill, and 3 warm-up iterations are executed to stabilize the
instruction cache and TLB state. The timed invocation immediately follows the
final warm-up, ensuring cold L2 but warm instruction paths.

\paragraph{Iteration-count extraction.}
The replay-based Phase~2 and Phase~4 iteration statistics used in the main text
were obtained separately from the timing experiment via a kernel-faithful
replay of the same control flow on the captured logits and previous-step
\texttt{preIdx} inputs, without perturbing the timed CUDA kernel path. This
replay covers 9 layers $\times$ 2{,}023 valid decode steps
(18{,}207 invocations) and records the Phase~2 loop count, done type, and
Phase~4 snap count per invocation. Because the replay reconstructs both the
Phase~2 bracket evolution and the Phase~4 candidate-buffer refinement from the
same logits and previous-step \texttt{preIdx}, it faithfully reproduces the
iteration statistics used in this analysis while avoiding additional in-kernel
counters.

\paragraph{Reproducibility.}
The entire experiment was repeated 3 times independently (from process startup
through all layers and steps). Per-layer average phase breakdowns show
${<}$0.3\% variation across runs. For example, L0 total latency was
34.3/34.4/34.3\,$\mu$s across the three runs, and L21 total was
23.4/23.3/23.4\,$\mu$s. Phase-level percentages were stable to ${\pm}$0.1
percentage points, confirming that the \texttt{clock64()}-based measurements
are highly reproducible under controlled conditions (locked SM clock, L2 flush,
warm-up).

\end{document}